\documentclass[useAMS,usenatbib]{mn2e}
\usepackage{amsmath}
\usepackage{graphicx}
\usepackage{natbib}  
\newcommand{\ha}{H$\alpha$}

\newcommand{\kms}{\,km\,s$^{-1}$}
\newcommand{\myr}{\,$M_{\sun}\,{\rm yr}^{-1}$}
\newcommand{\ro}{\,$R_{\sun}$}

\newcommand{\lo}{\,$L_{\sun}$}

\newcommand{\cmd}{\,cm$^{-2}$}
\newcommand{\cmt}{\,cm$^{-3}$}

\newcommand{\ecsa}{$\,\rm\,erg\,cm^{-2}\,s^{-1}\,\AA^{-1}$}
\def\I{\rm {\scriptsize I}}
\def\V{\rm {\scriptsize V}}
\title[Parameters of symbiotic nebulae from Thomson scattering]
      {Electron optical depths and temperatures of symbiotic 
       nebulae from Thomson scattering}

\author[M. Seker\'a\v{s} and A. Skopal]
       {M. Seker\'a\v{s}\thanks{E-mail: sekeras@ta3.sk (MS); 
        skopal@ta3.sk (AS)} and A. Skopal\\
        Astronomical Institute, Slovak Academy of Sciences, 
        059\,60 Tatransk\'a Lomnica, Slovakia}
\begin{document}

\date{Accepted 2012 August 24, Received 2012 July 23; 
      in original form 2012 May 23}

\pagerange{\pageref{firstpage}--\pageref{lastpage}} \pubyear{2012}

\maketitle

\label{firstpage}

\begin{abstract}
Symbiotic binaries are comprised of nebulae, whose densest portions 
have electron concentrations of $10^8 - 10^{12}$\cmt\ and 
extend to a few AU. They are optically thick enough to cause 
a measurable effect of the scattering of photons on free 
electrons. 
In this paper we introduce modelling the extended wings of 
strong emission lines by the electron scattering with the 
aim to determine the electron optical depth, $\tau_{\rm e}$, 
and temperature, $T_{\rm e}$, of symbiotic nebulae. 
We applied our profile-fitting analysis to the broad wings of 
the O\V\I\,1032,\,1038\,\AA\ doublet and He\I\I\ 1640\,\AA\ emission 
line, measured in the spectra of symbiotic stars AG~Dra, Z~And and 
V1016~Cyg. Synthetic profiles fit well the observed wings. 
By this way we determined $\tau_{\rm e}$ and $T_{\rm e}$ of 
the layer of electrons, throughout which the line photons are 
transferred. 
During quiescent phases, the mean 
  $\tau_{\rm e} = 0.056\pm 0.006$ and 
  $T_{\rm e} = 19\,200\pm 2\,300$\,K, 
while during active phases, mean quantities of both parameters 
increased to 
  $\tau_{\rm e} = 0.64\pm 0.11$ and 
  $T_{\rm e} = 32\,300\pm 2\,000$\,K. 
During quiescent phases, the faint electron-scattering wings 
are caused mainly by free electrons from/around the accretion 
disk and the ionized wind from the hot star with the total 
column density, $N_{\rm e} \la 10^{23}$\,\cmd.
During active phases, the large values of $\tau_{\rm e}$ are 
caused by a supplement of free electrons into the binary 
environment as a result of the enhanced wind from the hot star, 
which increases $N_{\rm e}$ to $\sim 10^{24}$\,\cmd. 
\end{abstract} 

%
%

\section{Introduction}

Symbiotic stars are long-period interacting binaries with orbital 
periods in the range of years. In their spectra we can recognize 
three main sources of radiation. The first one is represented 
by a giant star of spectral type (G-)K-M, the second one is 
a very hot ($T_{\rm h}\ga 10^5\,\rm K$) compact star, which is in 
most cases a white dwarf accreting from the wind of the giant. 
The third component of radiation is produced by a nebula, which 
represents the ionized fraction of the circumstellar material 
in the binary \citep[e.g.][ and references therein]
{boy67,stb,k86,nv87,cmm03,sk05}. 
As a result, the circumstellar environment of symbiotic binaries 
comprises energetic photons from the hot star with luminosities of 
$10^2 - 10^4$\,\lo\ \citep[e.g.][]{m91,greiner+97,sk05}, neutral 
particles produced by the cool giant at rates of a few times 
$10^{-7}$\,\myr\ \citep[e.g. STB,][]{mio02,sk05}, and ions 
and free electrons resulting from the processes of ionization. 
Symbiotic stars thus represent ideal objects for studying effects 
of Rayleigh, Raman and Thomson scattering. Raman and Rayleigh 
scattering results from interaction between the hot star photons 
and the neutral atoms in the giant's wind. They are important 
tools in mapping the ionization structure of symbiotic binaries 
\citep[e.g.][]{inv89,nsv89,schmid98,birriel04,lee09,lee12}. 
In contrast, the Thomson scattering of photons by free electrons 
acts within the ionized part of the symbiotic stars environment, 
and thus can diagnose the symbiotic nebula. In the spectrum, we can 
indicate this effect in the form of shallow, wide wings of 
the strongest emission lines, which photons are scattered by 
free electrons, and thus are Doppler shifted by their thermal 
motion to both the red and blue side of the line. 
The effect of this process is weak and wavelength independent, 
because of a very small and constant value of the Thomson 
cross-section, $\sigma_{\rm T} = 6.652\times 10^{-25}$\,cm$^{2}$. 
Nevertheless, the densest portions of symbiotic nebulae with 
electron concentrations of $\log(n_{\rm e}) \sim 8 - 12$ 
($n_{\rm e}$ in \cmt), could be optically thick enough to cause 
a measurable effect of the electron scattering. 
From this point of view, strong emission lines of highly ionized 
elements that are formed in the densest part of the ionized 
medium in a vicinity of the hot white dwarf, represent the best 
candidates. 

Originally, \cite{schmid+99} suggested that the broad wings of 
the O\V\I\ 1032 and 1038\,\AA\ resonance lines could be explained 
by scattering of the O\V\I\ photons by free electrons. 
\cite{young+05} successfully compared a model of the electron 
scattering wings with the O\V\I\ doublet in the AG~Dra spectrum. 
This process was also used by \cite{jung+lee} to model the broad 
\ha\ wings in the spectrum of the symbiotic star V1016~Cyg as 
an alternative to the Raman scattering of Ly$\beta$ photons on 
atomic hydrogen. \cite{sk+09} used the electron scattering of 
the O\V\I\,1032,\,1038\,\AA\ doublet and He\I\I\,1640\,\AA\ 
line in the AG~Dra spectrum to support the origin of the 
X-ray--UV flux anticorrelation, revealed by modelling the SED. 

In this paper we describe a simplified method for fitting 
the broad wings of intense emission lines by the 
electron scattering process (Sect.~3). 
We applied our profile-fitting procedure to the O\V\I\ 
resonance doublet and the He\I\I\,1640\,\AA\ line, observed 
in the spectra of symbiotic stars AG~Dra, Z~And and V1016~Cyg, 
with the aim to determine the electron optical depth, 
$\tau_{\rm e}$, and temperature, $T_{\rm e}$, of their 
nebulae. Sect.~4 presents the results and their discussion.  
Conclusions are found in Sects.~5. 

\section{Observations and data treatment}

For the purpose of this paper, we used extremely intense 
emission lines of the O\V\I\,1032,\,1038\,\AA\ doublet 
observed in the 
\textsl{FUSE} ({\em Far Ultraviolet Spectroscopic Explorer}), 
\textsl{BEFS} ({\em Berkeley Extreme and Far-UV Spectrometer}) 
and 
\textsl{TUES} ({\em T\"{u}bingen Ultraviolet Echelle Spectrograph}) 
spectra of AG~Dra, Z~And and V1016~Cyg. We used also the strong
emission line He\I\I\,1640\,\AA\ in the high-resolution 
\textsl{IUE} ({\em International Ultraviolet Explorer}) 
spectra of AG~Dra. 
The spectra were obtained from the satellite archives with the 
aid of the Multimission Archive at the Space Telescope Science 
Institute (MAST). They are summarized in Table~1. 

The \textsl{FUSE} spectra were processed by the calibration 
pipeline version 3.0.7, 3.0.8 and 2.4.1. We used the calibrated 
time-tag observations ({\small TTAG} photon collecting mode). 
Before adding the flux from all exposures we applied an
appropriate wavelength shift relative to one so to get 
the best overlapping of the absorption features. Then we
co-added spectra of individual exposures and weighted them
according to their exposure times. The wavelength scale of 
the spectra was calibrated with the aid of the interstellar 
absorption lines \citep[e.g.][]{re84}. Accuracy of such 
calibration is of $\pm$\,0.05\,\AA. Finally, we binned 
the resulting spectrum within 0.025\,\AA. 
%
\begin{table}
\begin{center}
\caption{Log of spectroscopic observations.       
         }
\begin{tabular}{ccccc}
\hline
\hline
   Date      &  Julian date  & Orb.         &  Obs.      & Line$^{b})$ \\
(dd/mm/yyyy) & - 2\,400\,000 & phase$^{a})$ & Satellite  &      \\
\hline\\[-3mm]
\multicolumn{5}{c}{AG~Dra} \\
\hline
 11/12/1981  &44949.5 &0.38  & \textsl{IUE}   & He\I\I\  \\
 18/09/1993  &49248.5 &0.22  & \textsl{BEFS}  & O\V\I\   \\
 29/06/1994  &49532.5 &0.74  & \textsl{IUE}   & He\I\I\  \\
 09/07/1994  &49542.5 &0.75  & \textsl{IUE}   & He\I\I\  \\
 12/07/1994  &49545.5 &0.76  & \textsl{IUE}   & He\I\I\  \\
 28/07/1994  &49561.5 &0.79  & \textsl{IUE}   & He\I\I\  \\
 17/09/1994  &49612.5 &0.88  & \textsl{IUE}   & He\I\I\  \\
 28/07/1995  &49927.5 &0.46  & \textsl{IUE}   & He\I\I\  \\
 14/02/1996  &50127.5 &0.82  & \textsl{IUE}   & He\I\I\  \\
 22/11/1996  &50409.5 &0.33  & \textsl{TUES}  & O\V\I\   \\
 16/03/2000  &51619.5 &0.54  & \textsl{FUSE}  & O\V\I\   \\
 25/04/2001  &52024.5 &0.28  & \textsl{FUSE}  & O\V\I\   \\
 14/11/2003  &52957.5 &0.98  & \textsl{FUSE}  & O\V\I\   \\
 24/06/2004  &53180.5 &0.38  & \textsl{FUSE}  & O\V\I\   \\
 25/12/2004  &53364.5 &0.72  & \textsl{FUSE}  & O\V\I\   \\
 15/03/2007  &54174.5 &0.20  & \textsl{FUSE}  & O\V\I\   \\
\hline\\[-3mm]
\multicolumn{5}{c}{Z~And} \\
\hline
 05/07/2002  &52460.5 &0.90  & \textsl{FUSE}  & O\V\I\  \\
 04/08/2003  &52855.5 &0.42  & \textsl{FUSE}  & O\V\I\  \\            
\hline\\[-3mm]
\multicolumn{5}{c}{V1016~Cyg} \\
\hline
 10/08/2000  &51766.5 & ? & \textsl{FUSE}  & O\V\I\  \\
\hline
\end{tabular}
\end{center}
$^{a})$ - According to \cite{fe00}, 
$^{b})$ - He\I\I\,1640\,\AA, O\V\I\ 1032,\,1038\,\AA\ doublet, 
\end{table}

All the spectra were corrected for heliocentric velocity 
including that of the satellite, and dereddened with 
$E_{\rm B-V}$ = 0.30 \citep[Z~And,][]{m91}, 0.28 
\citep[V1016~Cyg,][]{ns8} and 0.08 for AG Dra \citep[][]{bi00}, 
using the extinction curve of \cite{ccm89}. 

\section{The model}

\subsection{Assumptions and simplifications}

Thomson scattering represents a special case of the scattering 
of a photon off a free electron, which is fully elastic (i.e. 
the photon energy does not change). In the real case, the photon 
always transfers some of its energy to the electron, which 
shifts its wavelength by $\sim$+0.024~\AA, the so-called Compton 
wavelength of the electron. However, the Doppler effect arising 
from the thermal motion of free electrons leads to significantly 
larger shifts to both sides of the spectrum. Therefore, the 
elastic Thomson scattering is a good approximation in studying 
scattering of low-energy photons ($h\nu << m_{\rm e}c^2$) off 
non-relativistic electrons 
\citep[see e.g.][ in detail]{rosswog07}. 

To model the effect of the electron scattering we have to know 
how the scattered photons are redistributed in frequencies and 
directions. It was shown by \cite{hum+mih67} that it is convenient 
and sufficiently accurate to regard the radiation field, from which 
scattering occurs, as isotropic, so that the direction can be 
averaged out. Therefore, for the sake of simplicity, we consider 
isotropic scattering with Maxwellian distribution of electron 
velocities. Under these assumptions the redistribution function 
can be expressed in a form \citep[][]{mihalas70}, 
\begin {equation} 
 R_{\rm e}\left(\nu,\nu^{'}\right)=
   \frac{1}{w}
   \left(\frac{e^{-{\left|\frac{\nu-\nu^{'}}{2w}\right|}^{2}}}
    {\sqrt{\pi}}\right)-\left
   |\frac{\nu-\nu^{'}}{2w}\right
   |{\rm erfc}\left|\frac{\nu-\nu^{'}}{2w}\right|,
\end {equation}
where $\nu$ and $\nu^{'}$ are frequencies of radiation before and 
after the scattering and $w$ is the electron Doppler width, 
\begin {equation} 
  w = \frac{\nu_{0}}{c}\sqrt{\frac{2k T_{\rm e}}{m_{\rm e}}}, 
\end {equation}
and the complementary error function $\rm{erfc(x)}$ is defined as 
\begin {equation} 
  {\rm erfc}\left(x\right)=
      \frac{2}{\sqrt{\pi}}\int^{\infty}_{x}e^{-z^{2}}dz.
\end {equation}

To calculate the electron-scattering wings profiles, we
adopted a simplified scheme of \cite{munch50}, which assumes 
that a plane-parallel layer of free electrons of the optical 
thickness $\tau_{\rm e}$ and the temperature $T_{\rm e}$ is 
irradiated by the line photons, and that the electrons are 
segregated from the other opacity sources, which implies no 
change in the equivalent width of the line. 
In our case, this assumption corresponds to modelling the wings 
of highly ionized O\V\I\ and He\I\I\ lines, which are formed 
within the O$^{+5}$ and/or He$^{+}$ zone close to the hot white 
dwarf photosphere, and the layer of free electrons, where the 
Thomson scattering arises, is located above the line formation 
region. 
%

According to the radiative transfer equation, assuming that the 
electron scattering is the only process attenuating the original 
line photons, the observed line flux can be expressed as 
%
%
\begin{equation}
  F^{\rm obs} = F_0\,e^{-\tau_{\rm e}},
\label{eq:fobs}
\end{equation}
where $F_0$ is the line flux before scattering, 
$\tau_{\rm e} = \sigma_{\rm T} N_{\rm e}$ is the electron 
optical depth and $N_{\rm e}$ is the column density of free 
electrons along the line of sight. As the scattered fraction 
of the original flux is redistributed into the line wings so 
that the equivalent width before and after the scattering is 
constant, 
$F_{\rm wing} = F_0 - F^{\rm obs} = F_0 (1 - e^{-\tau_{\rm e}})$. 
Then the line profile after it emerges from the layer of 
scattering electrons may be approximated by 
\citep[see also][]{castor+70} 
\begin {equation}
 \Psi(x) = e^{-\tau_{\rm e}} \Phi (x) + 
           (1 - e^{-\tau_{\rm e}})\! \int\limits_{\infty}^{-\infty}
           \! \Phi(x^{'})R_{e}(x^{'},x){\rm d}x^{'},
\label{eq:psi}
\end {equation}
where $\Phi(x)$ is the incident line profile, 
$R_{\rm e}(x^{'},x)$ is the redistribution function for 
Thomson scattering (Eq.~(1)) and $x$ or $x^{'}$ is a frequency 
displacement from the line center in units of electron Doppler 
width before and after the scattering, respectively. 
So, the first right-side term 
of Eq.~\eqref{eq:psi} represents the original flux at $x$ 
attenuated by the scattering (Eq.~\eqref{eq:fobs}), and its 
scattered, $(1 - e^{-\tau_{\rm e}})$, fraction is redistributed 
in the wings (the second term). Note that the scattered profile 
used previously by e.g. \cite{castor+70} represents a special 
case of Eq.~\eqref{eq:psi} for $\tau_{\rm e} \ll 1$. 

To fit the theoretical profile (\ref{eq:psi}) to observations, 
means to determine its variables, $\tau_{\rm e}$, $T_{\rm e}$, 
and those of the incident profile $\Phi(x)$. 
%
%
\begin{figure}
\centering
\begin{center}
\resizebox{\hsize}{!}{\includegraphics[angle=-90]{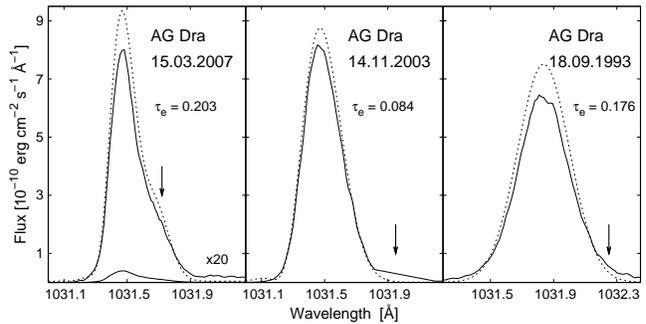}}
\end{center}
\caption[]{
Examples of the observed (solid lines) and the incident 
(dotted lines) profiles of the O\V\I\,1032\,\AA\ line. 
Arrows denote the red-side emission bumps. 
}
\label{fig:in}
\end{figure}
%
%
%
%

\subsection{The incident profile} 

First, we estimated the continuum level from a large wavelength 
interval around the O\V\I\ lines by a linear fit to the noise 
in the spectrum. In the case of the He\I\I\ 1640\,\AA\ line, 
the continuum level was estimated with the aid of 
the corresponding low-resolution spectrum. 
%
Second, we approximated the incident profile $\Phi(\lambda)$ 
in the model (\ref{eq:psi}) as follows. 
The He\I\I\ 1640\,\AA\ line was possible to fit with a single 
Gauss curve. The fit of the observed emission core provided 
the first estimate of its position, width and the height. 
The O\V\I\,1032\,\AA\ line is in most cases asymmetric with 
respect to its reference wavelength. Its blue emission wing 
is steeper than the red one, being cut by an absorption 
component (see Fig.~\ref{fig:in}). It is probably caused 
by the scattering in the line at the close vicinity of the 
white dwarf within the densest part of the wind moving 
to the observer. This absorption can be more pronounced 
during active phases, when a higher mass-loss rate is observed 
\citep[][]{sk06}. The observation of AG~Dra from 2007 May 15, 
made during the 2006-08 active phase, is consistent with this 
view (see the left panel of Fig.~\ref{fig:in}). 
The scattering in the line operates within the line formation 
region. Therefore, we take the incident profile of the 
O\V\I\,1032\,\AA\ line as a sum of emission and absorption 
Gaussians. 
The incident emission of the O\V\I\ 1038\,\AA\ line was not 
possible to reconstruct by a direct fitting of its observed 
remainder, because of a strong influence by the absorption 
of the interstellar molecular hydrogen, $H_{2}$ 
\citep[e.g.][]{schmid+99}. Therefore, we reconstructed the 
incident O\V\I\ 1038\,\AA\ profile with the aid of the 
theoretical ratio of the doublet lines with the assumption 
that they have the same width (see below). 

Based on the above mentioned observational properties of 
the O\V\I\,1032,\,1038\,\AA\ doublet, we reconstructed its 
incident profile by a superposition of three Gaussians, 
\begin{equation}
 \Phi(\lambda) = \sum_{\rm n=1}^{\rm 3}
       I_{\rm n}\exp\left[{-\frac{1}{2}\left(\frac{\lambda-
       \lambda_{\rm n}}{\sigma_{\rm n}}\right)^{2}}\right], 
\label{eq:fi}
\end{equation}
where indices n = 1 and 3 denote the Gaussians of the emission 
cores at $\lambda_{1}\sim 1032$\,\AA\ and 
$\lambda_{3}\sim 1038$\,\AA. The second curve (n = 2) represents 
the absorption component, which cut the blue side of 
the 1032\,\AA\ emission line (Fig.~\ref{fig:in}). 
However, this rather strong absorption did not reproduce the 
narrow one seen in the scale of Figs.~\ref{fig:agmod} and 
\ref{fig:zmod} as a P-Cyg component in the 1032\,\AA\ line, 
but not recognizable in the scale of Fig.~\ref{fig:in}. 
Only for the V1016~Cyg spectrum, due to a symmetrical emission 
core of the 1032\,\AA line, it was possible to fit this narrow 
P-Cyg absorption by the $\I_{2}$ component (see the right panel 
of Fig.~\ref{fig:zmod}). Irrespectively of its origin, we 
neglected its influece to the profile. 

The effect of the interstellar $H_{2}$ absorption to the 
O\V\I\,1038\,\AA\ line profile was significant in the spectra 
of Z~And and V1016~Cyg, because of a large amount of interstellar 
matter on the line of sight to these objects 
($E_{\rm B-V} \sim 0.3$). As a result, the observed intensity 
ratio $\I_{1}/\I_{3}$ was $\approx 7$ for Z~And and $\approx 15$ 
for V1016~Cyg. 
Therefore, we reconstructed the original O\V\I\,1038\,\AA\ 
line adopting the theoretical ratio $\I_{1}/\I_{3} = 2$, 
$\sigma_3 = \sigma_1$ and $\lambda_3$ so to fit the emission 
remainder of the line. 
The absorption effect from the $H_{2}$ molecules was apparently 
fainter on the AG~Dra spectra ($E_{\rm B-V} \sim 0.08$, 
$\I_{1}/\I_{3}\approx 2$), and thus the estimate of the initial 
O\V\I\,1038\,\AA\ line profile was more trustworthy. 

Some spectra show a noticeable additional emission features at 
the red side of the 1032\,\AA\ and 1038\,\AA\ line cores up to 
$\sim 10$\% of their height. 
We did not investigate the origin of these features and did not 
take them into account in the fitting procedure. However, in the 
AG~Dra spectrum from 2007 March, the unknown emission features 
were rather strong. Therefore, we included them to the incident 
O\V\I\ 1032\,\AA,\,1038\,\AA\ line profiles as an additional 
Gaussians (see Fig.~\ref{fig:in}). 

\subsection{Profile-fitting analysis}

First, we selected the flux-points of the observed profile(s), 
$F^{\rm obs}(\lambda_{\rm i})$, for fitting with the function 
(\ref{eq:psi}). They were selected from the observed profiles 
by omitting some artificial (sharp) emission/absorption features 
and the depression around the Ly${\beta}$ line, caused by the 
Rayleigh scattering (see Fig.~\ref{fig:agmod}). 
%
To find the best solution, we calculated a grid of models 
for reasonable ranges of the fitting parameters, $I_{\rm n}$, 
$\sigma_{\rm n}$, $\lambda_{\rm n}$ for the original profile 
$\Phi(x)$ and $\tau_{\rm e}$, $T_{\rm e}$ for the scattered 
wings. In the case of the O\V\I\ 1032\,\AA,\,1038\,\AA\ doublet, 
the parameters $I_3$, $\sigma_3$, $\lambda_3$ of its 1038\,\AA\ 
component were estimated according to the properties of both 
the doublet lines, as described in Sect.~3.2. 
The grid of models was prepared with steps 
$\Delta\tau_{\rm e} = 0.0005$, 
$\Delta T_{\rm e} = 100$\,K, 
$\Delta I_{\rm n} = I_{\rm n}/100$, 
$\Delta\sigma_{\rm n} = 0.001$ 
and 
$\Delta\lambda_{\rm n} = 0.001$\,\AA. 
Finally, we selected the model corresponding to a minimum of 
the function 
\begin{equation} 
  \chi^{2}_{\rm red} =
  \frac{1}{\aleph}\sum_{\rm i=1}^{\rm N}
               \left[\frac{F^{\rm obs}(\lambda_{\rm i})-
                     \Psi (\lambda_{\rm i})}
               {\Delta F^{\rm obs}(\lambda_{\rm i})}\right]^{2},
\label{eq:chi2}
\end{equation}
where $F^{\rm obs}(\lambda_{\rm i})$ are the observed fluxes of 
the profile, $N$ is their number ($\approx 200$), $\aleph$ is 
the number of d.o.f., $\Delta F^{\rm obs}(\lambda_{\rm i})$ are 
their errors and $\Psi(\lambda_{\rm i})$ are theoretical fluxes. 

Errors in the selected flux points were around of 10--15\% 
of the line wings. Based on them we determined uncertainties 
in $T_{\rm e}$ and $\tau_{\rm e}$ for individual spectra. 
To obtain a rough estimate of the corresponding range of 
$T_{\rm e}$ we fixed other fitting parameters and varied 
$T_{\rm e}$ so to fit fluxes $F^{\rm obs}(\lambda_{\rm i})
\pm \Delta F^{\rm obs}(\lambda_{\rm i})$. Similarly we 
proceeded to estimate the limits for $\tau_{\rm e}$. 
Such the estimated ranges of fitting parameters can be as 
large as $\sim 50$\% of the best model value and are 
asymmetrically placed with respect to it (see Table~2). 
This is a result of the non-uniformly distributed fluxes 
determining the profile of electron-scattered wings (see 
the beginning of Sect.~3.3.). 
In the case of the AG~Dra spectrum from 14/02/1996, the 
relatively small uncertainties in $T_{\rm e}$ and 
$\tau_{\rm e}$, but the large value of $\chi^2_{\rm red}$ 
probably reflect underestimated flux errors, 
$\Delta F^{\rm obs}(\lambda_{\rm i})$. 
Large flux uncertainties on the spectrum from 22/11/1996 
and poorly defined continuum level did not allow us 
to estimate the upper limit of $T_{\rm e}$ 
(see Fig.~\ref{fig:agmod}). 

\section{Results and Discussion}

The resulting fits are shown in Fig.~\ref{fig:agmod} and 
\ref{fig:zmod} and corresponding parameters are in Table~2. 
Significant changes in $\tau_{\rm e}$ and $T_{\rm e}$ 
reflect different properties of the symbiotic nebula during 
different levels of the activity. 
During quiescent phases, the mean 
   $T_{\rm e} = 19\,200 \pm 2\,300$\,K, 
   $\tau_{\rm e} = 0.056 \pm 0.006$, 
while during active phases 
   $T_{\rm e} = 32\,300 \pm 2\,000$\,K, 
   $\tau_{\rm e} = 0.64 \pm 0.11$, respectively. 
Here the uncertainties represent rms errors of the average 
values. 

The average quantities of $T_{\rm e}$ and $\tau_{\rm e}$ 
as derived from quiescent and active phases 
agree well with those found independently by modelling the SED 
\citep[e.g.][]{sk05,sk+09}. More than one order of magnitude 
difference in $\tau_{\rm e}$ (i.e. in $N_{\rm e}$) 
reflect a significant increase of free electrons on the line 
of sight in the direction to the hot star during active 
phases. 
The origin of this change is shortly discussed in Sects.~4.2. 
and 4.3. 

\subsection{Application to selected symbiotic stars}

\subsubsection{AG~Dra}
AG~Dra is a yellow symbiotic star comprising a cool giant 
of a K2\,III spectral type \citep{ms99}. It is located 
at a high galactic latitude of $41^{\circ}$, which implies that 
its spectrum is less affected by the interstellar matter. 
As a result, the line ratio $I$(1038\,\AA)/$I$(1032\,\AA) was 
close to its theoretical value of 0.5, which made the modelling 
of the O\V\I\ doublet more trustworthy (Sect.~3.2). 

\begin{figure*}
\centering
\begin{center}
\resizebox{\hsize}{!}{\includegraphics[angle=-90]{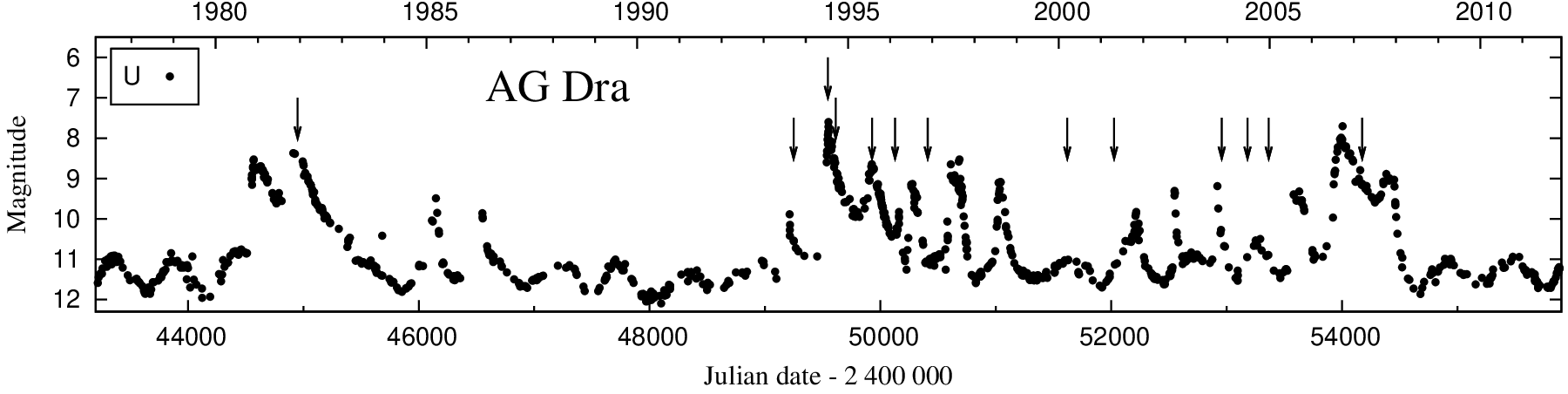}}
\resizebox{\hsize}{!}{\includegraphics[angle=-90]{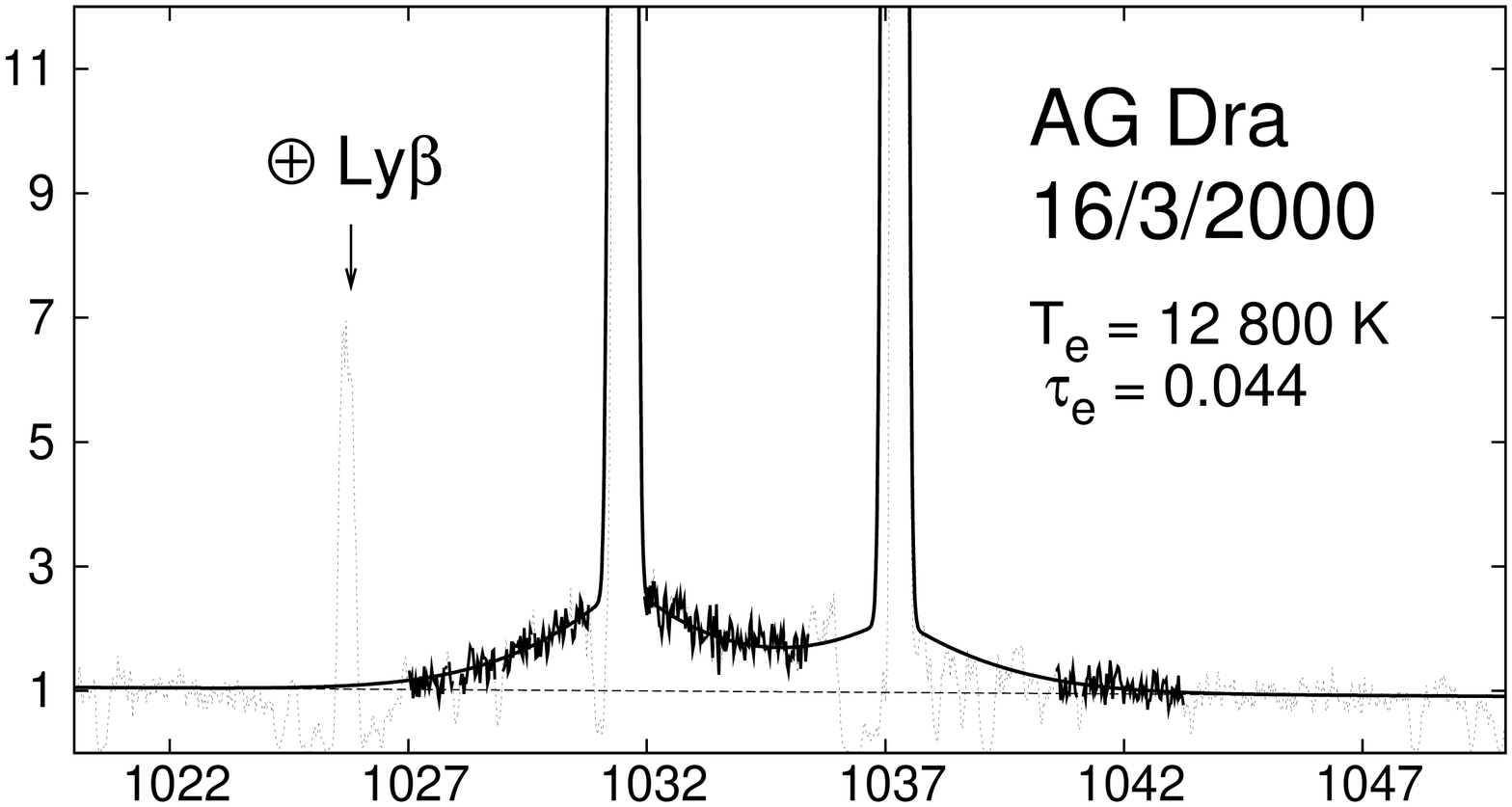}
                      \includegraphics[angle=-90]{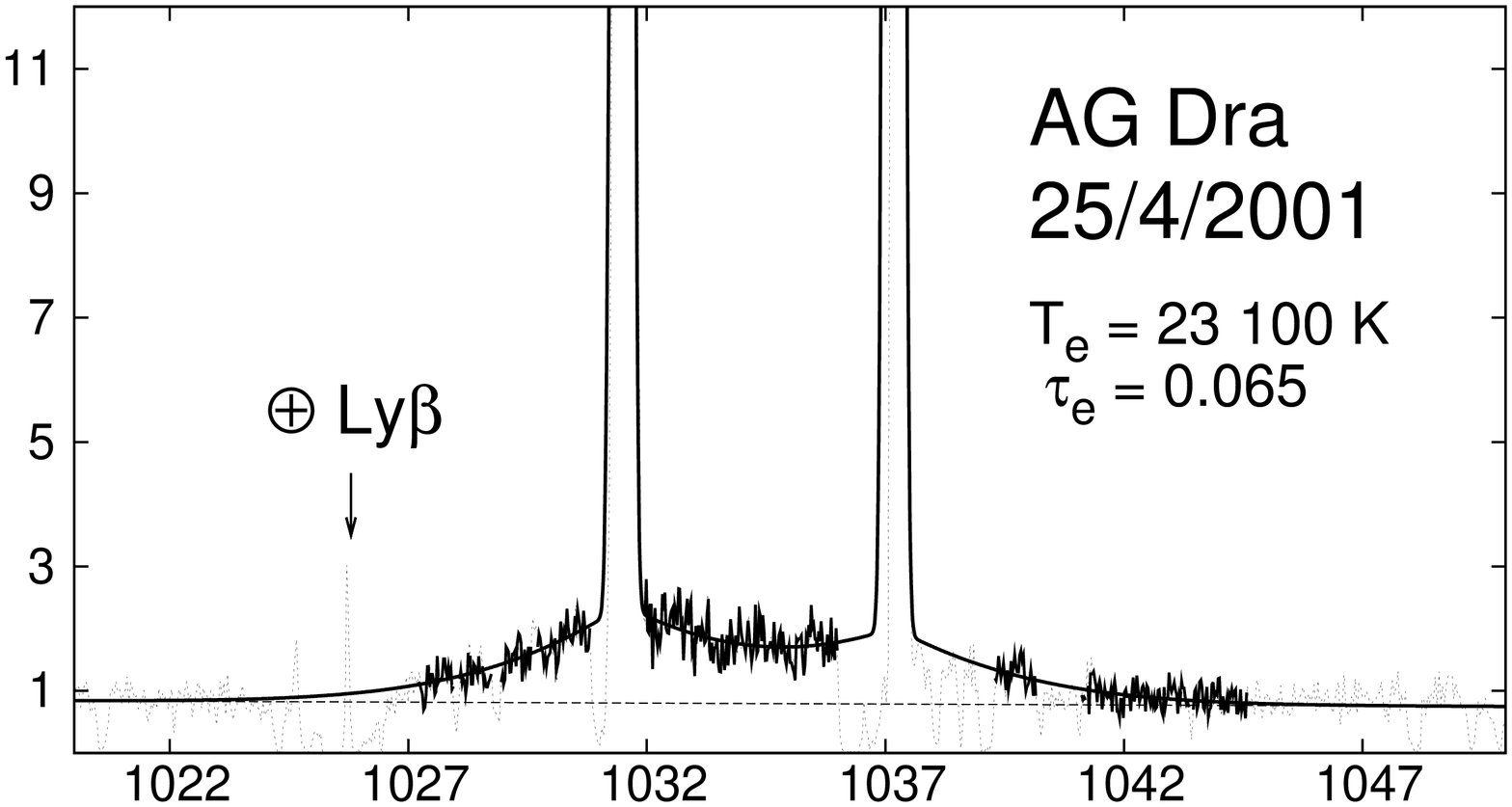}                      
                      \includegraphics[angle=-90]{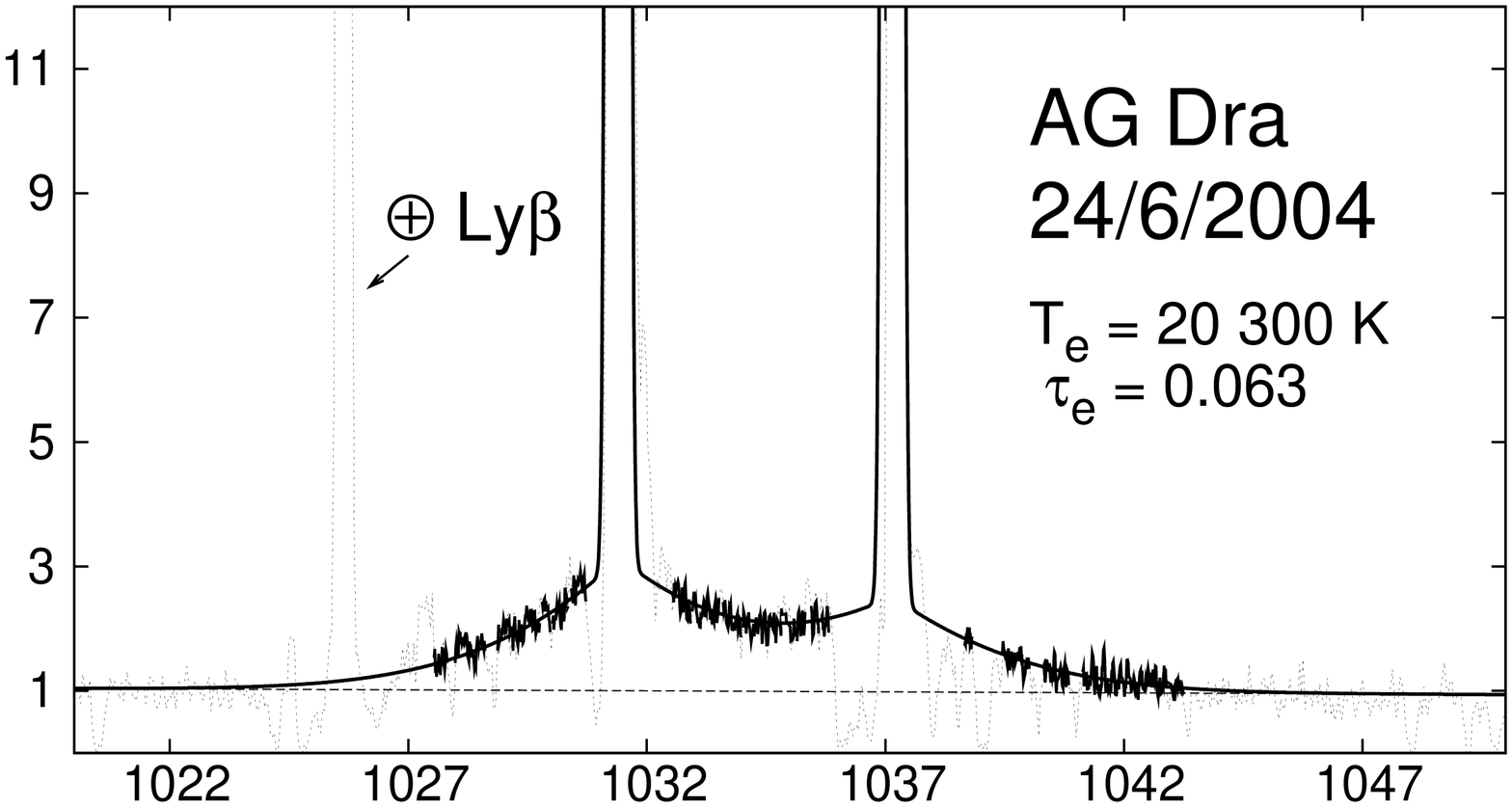}}
\resizebox{\hsize}{!}{\includegraphics[angle=-90]{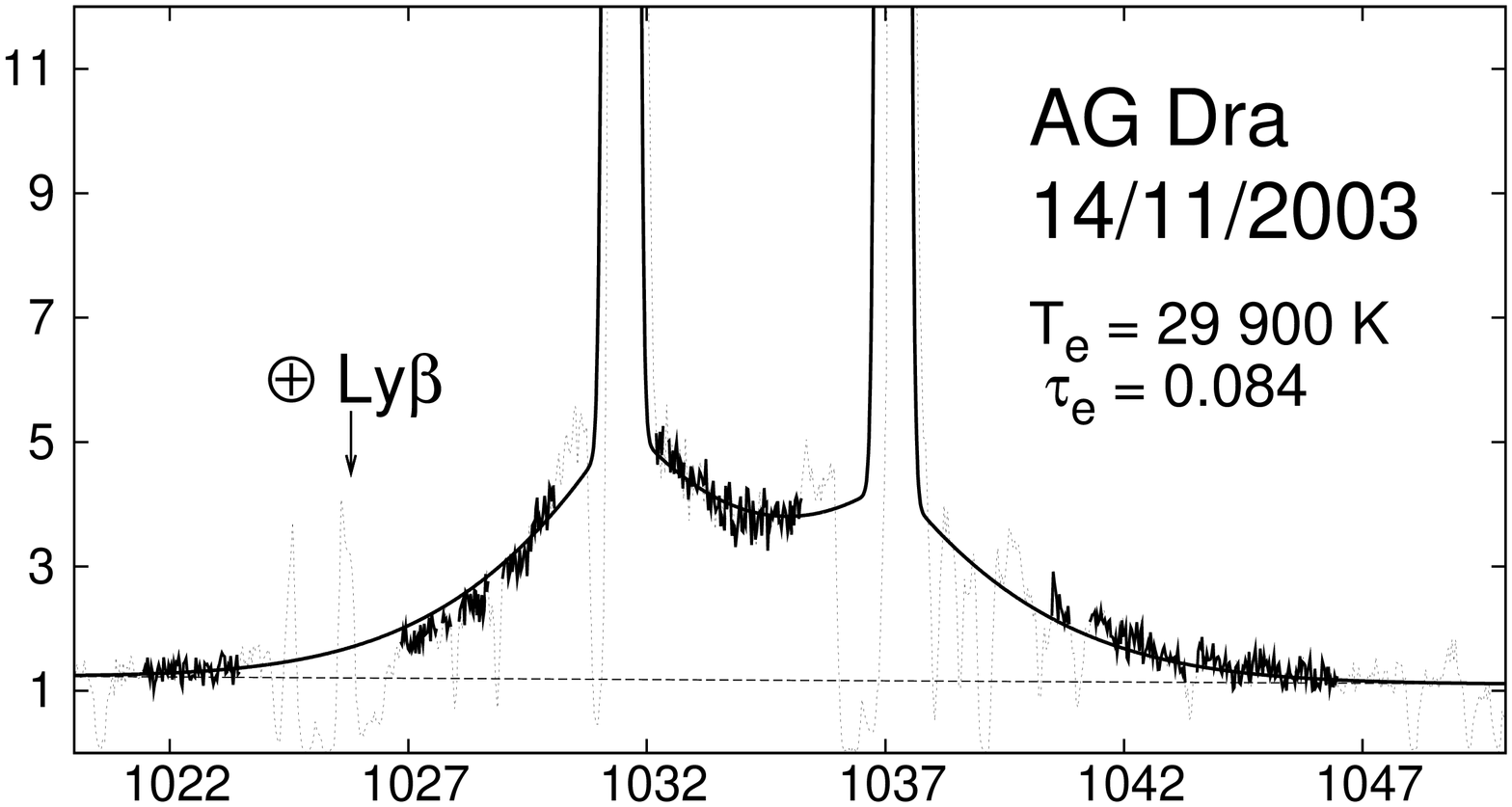}
                      \includegraphics[angle=-90]{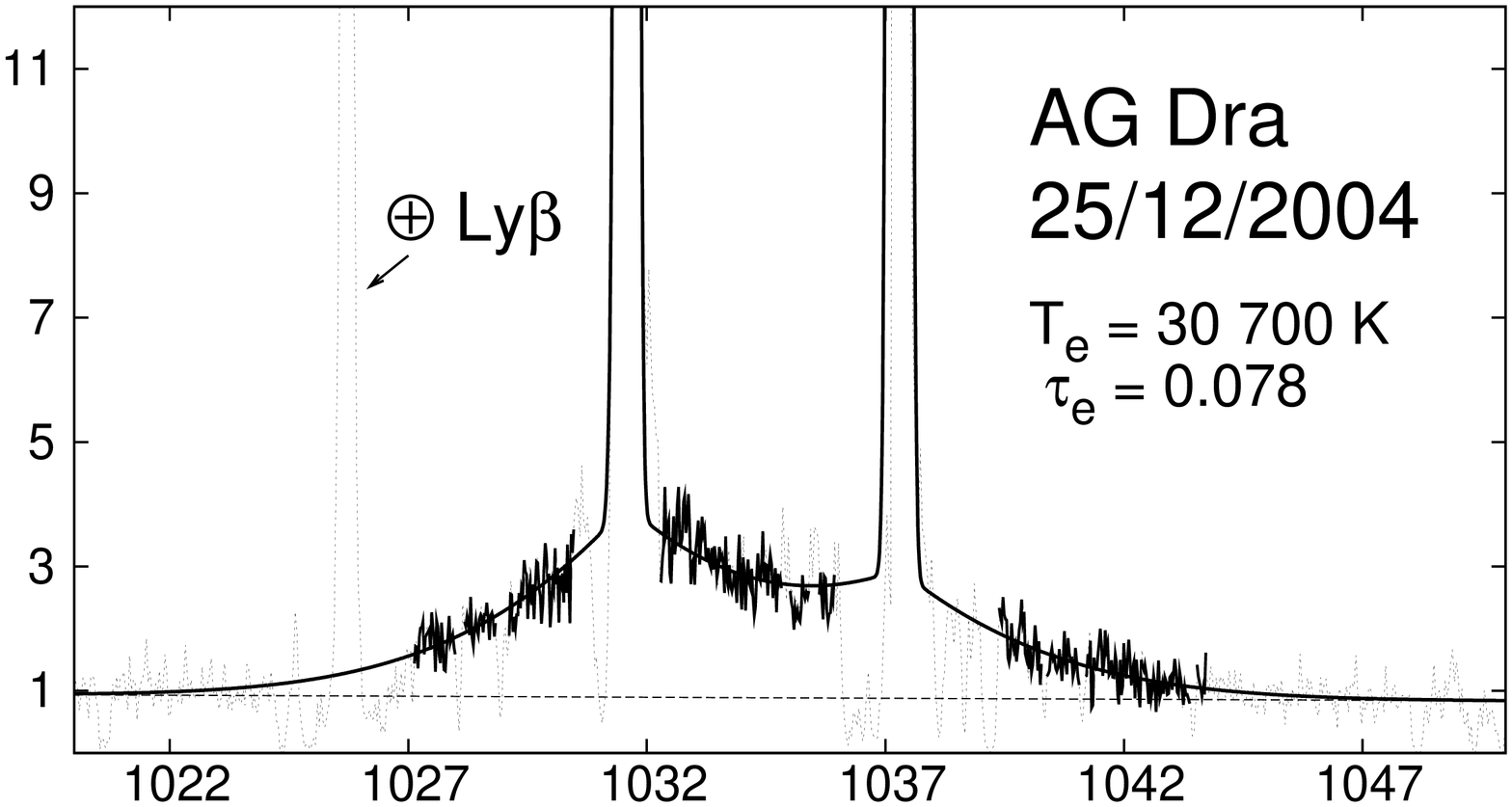}
                      \includegraphics[angle=-90]{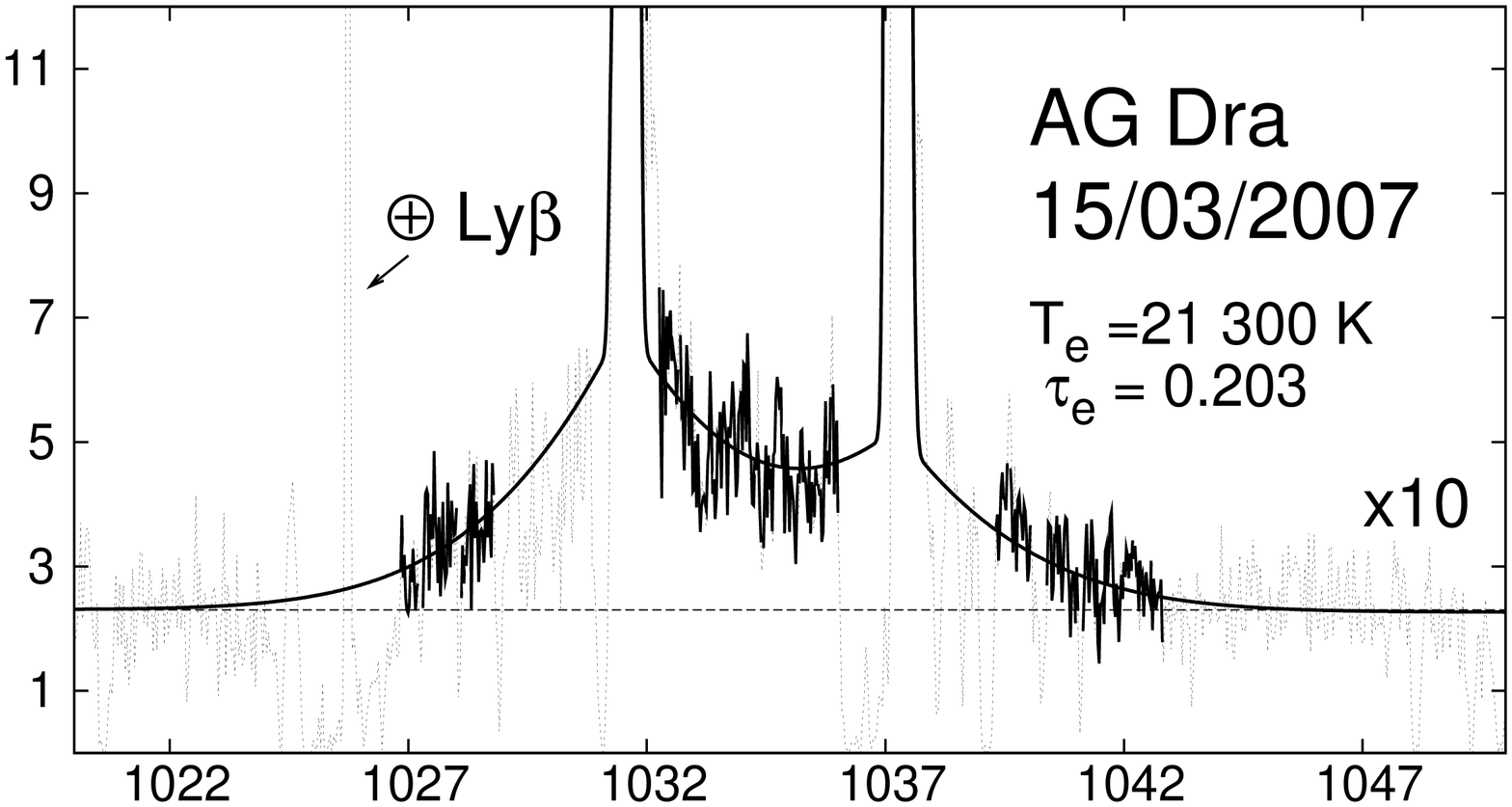}}
\resizebox{\hsize}{!}{\includegraphics[angle=-90]{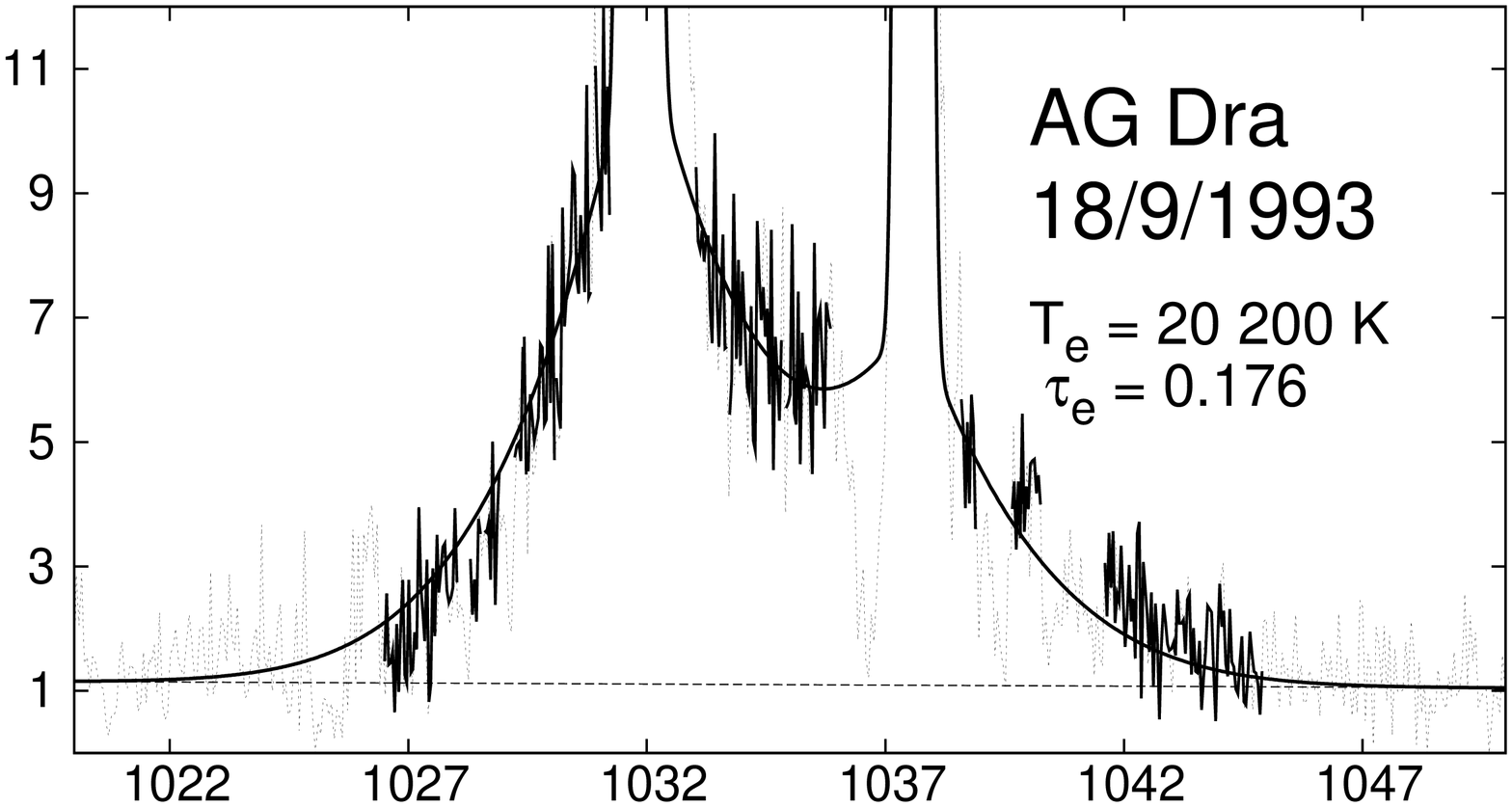}
                      \includegraphics[angle=-90]{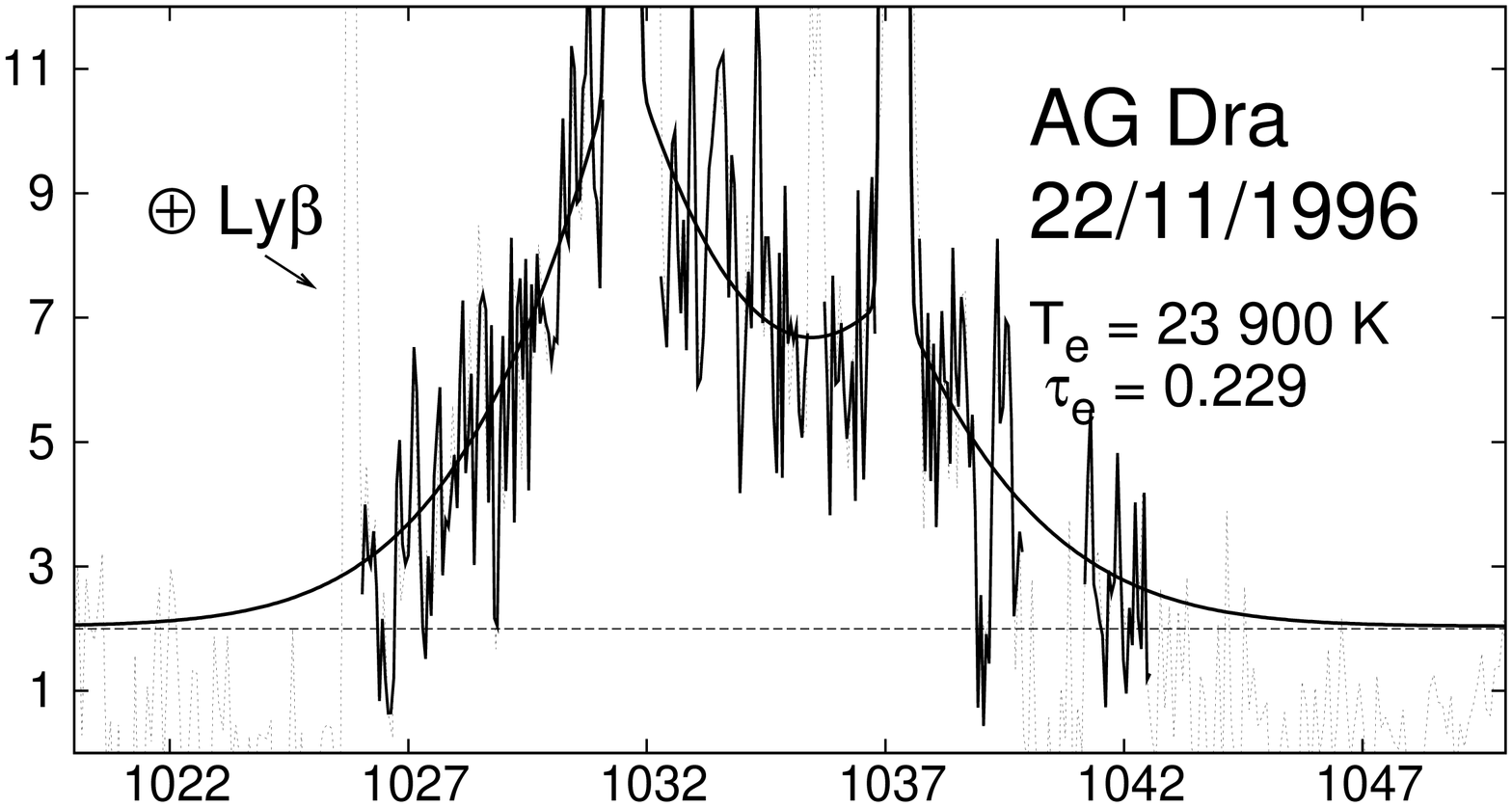}                   
                      \includegraphics[angle=-90]{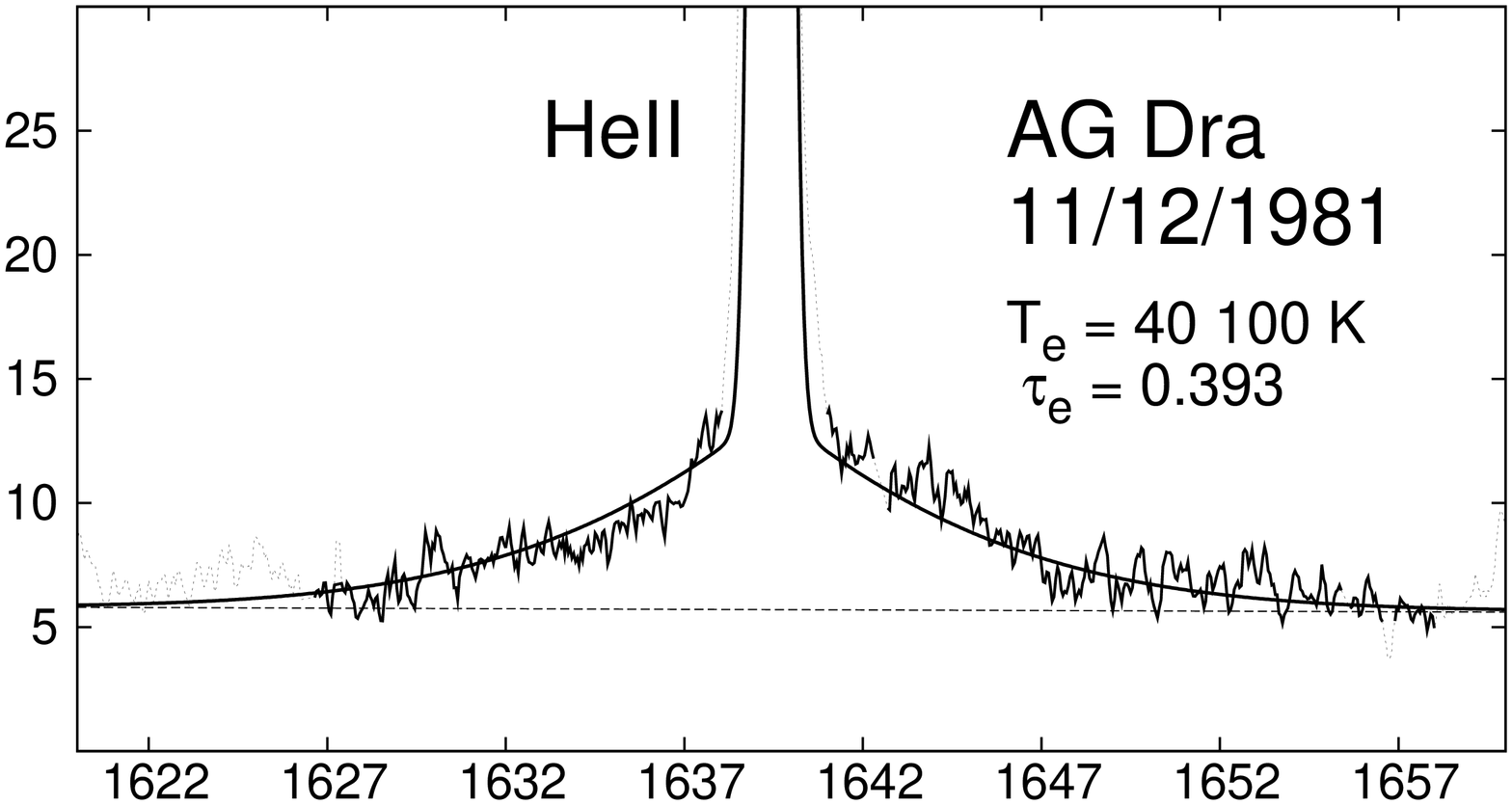}}
\resizebox{\hsize}{!}{\includegraphics[angle=-90]{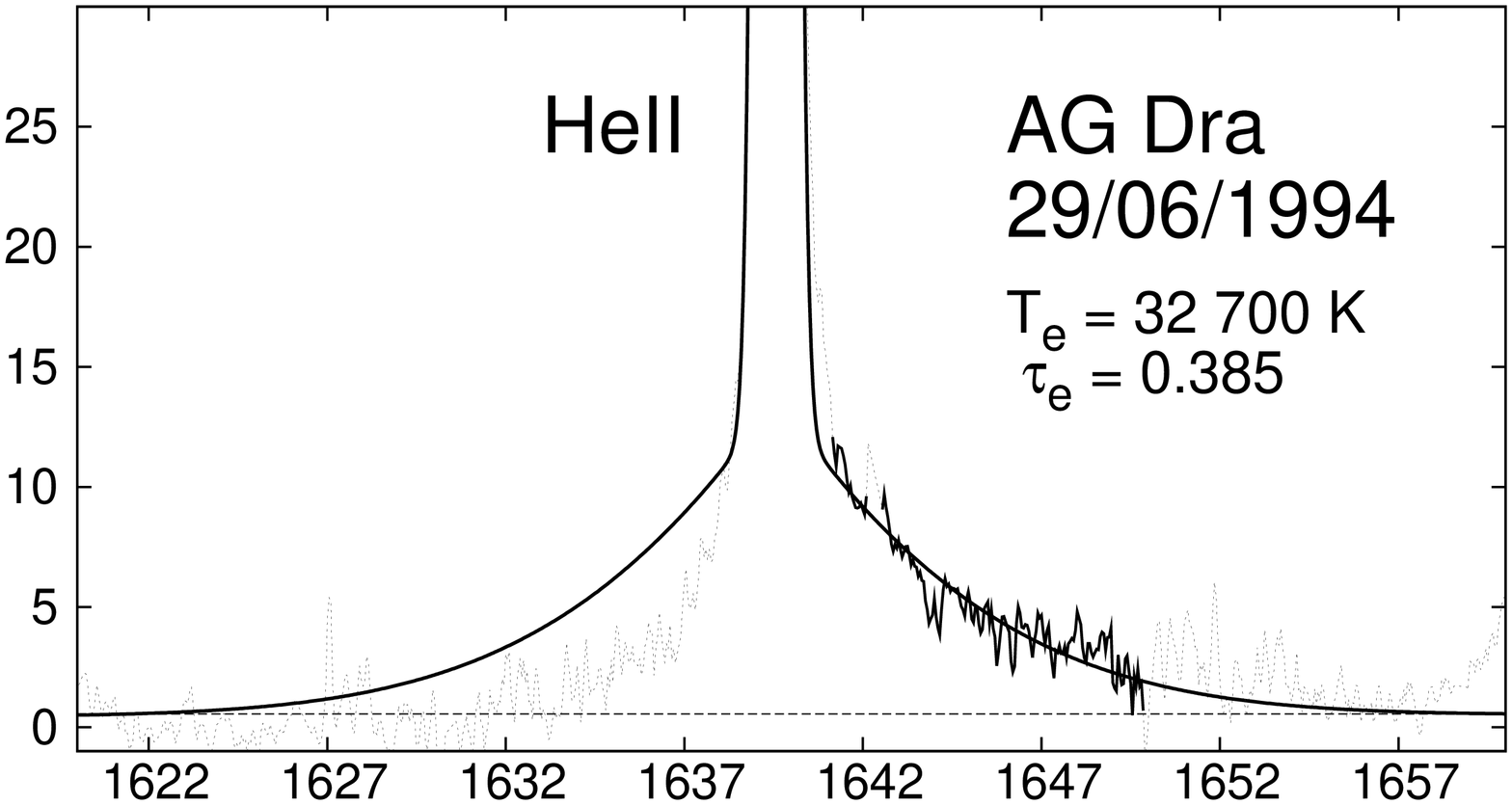}
                      \includegraphics[angle=-90]{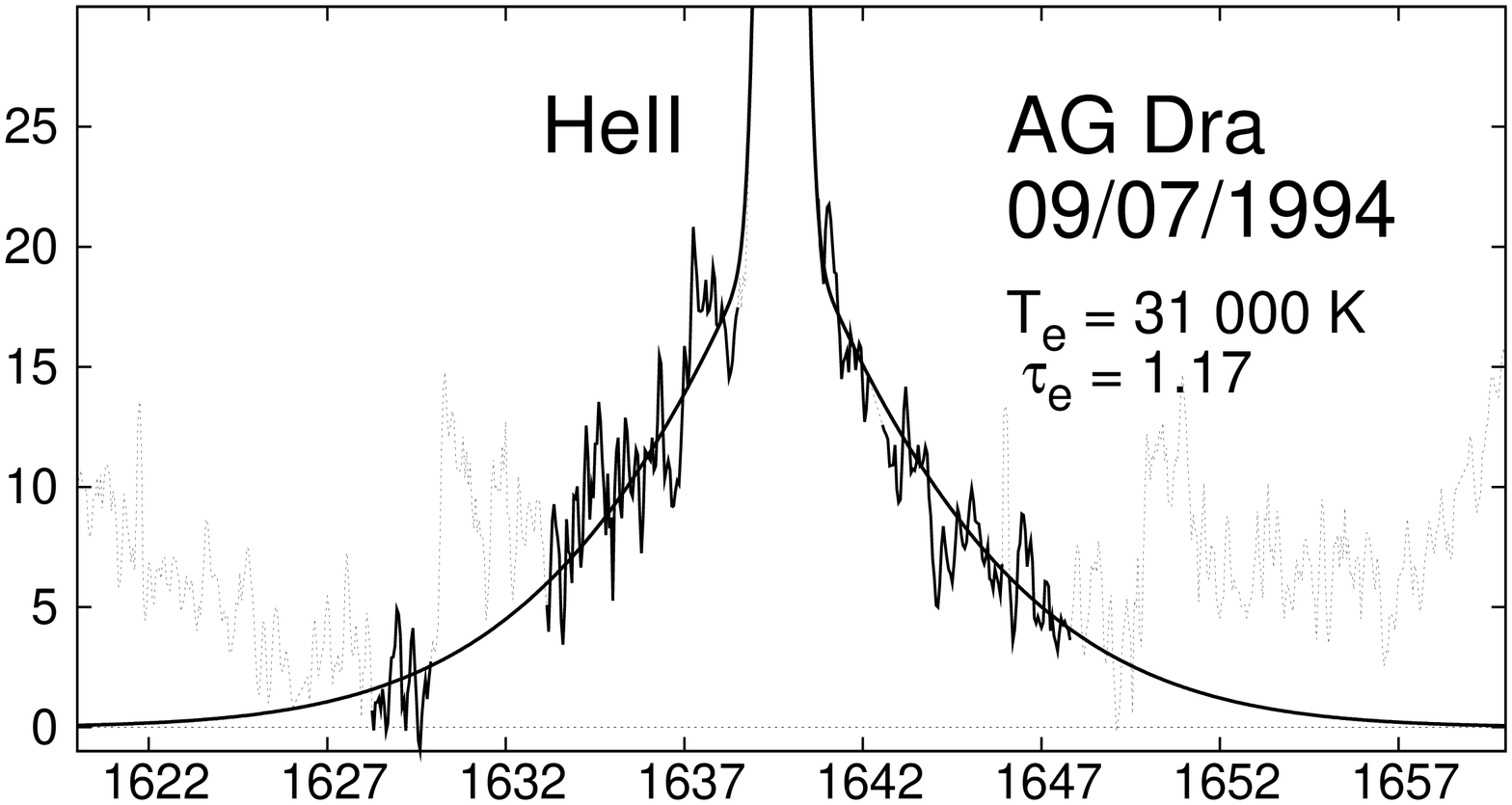}
                      \includegraphics[angle=-90]{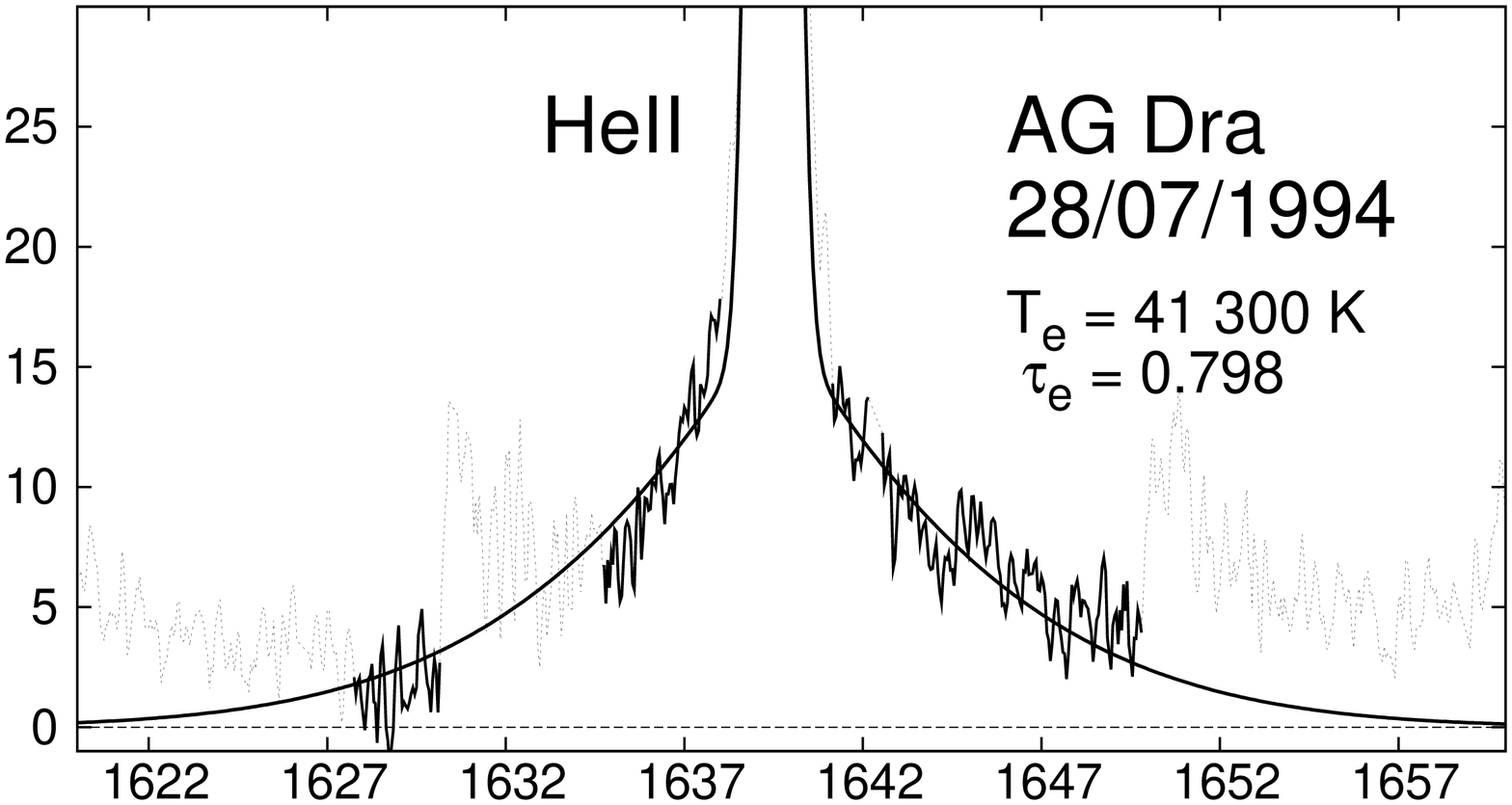}}
\resizebox{\hsize}{!}{\includegraphics[angle=-90]{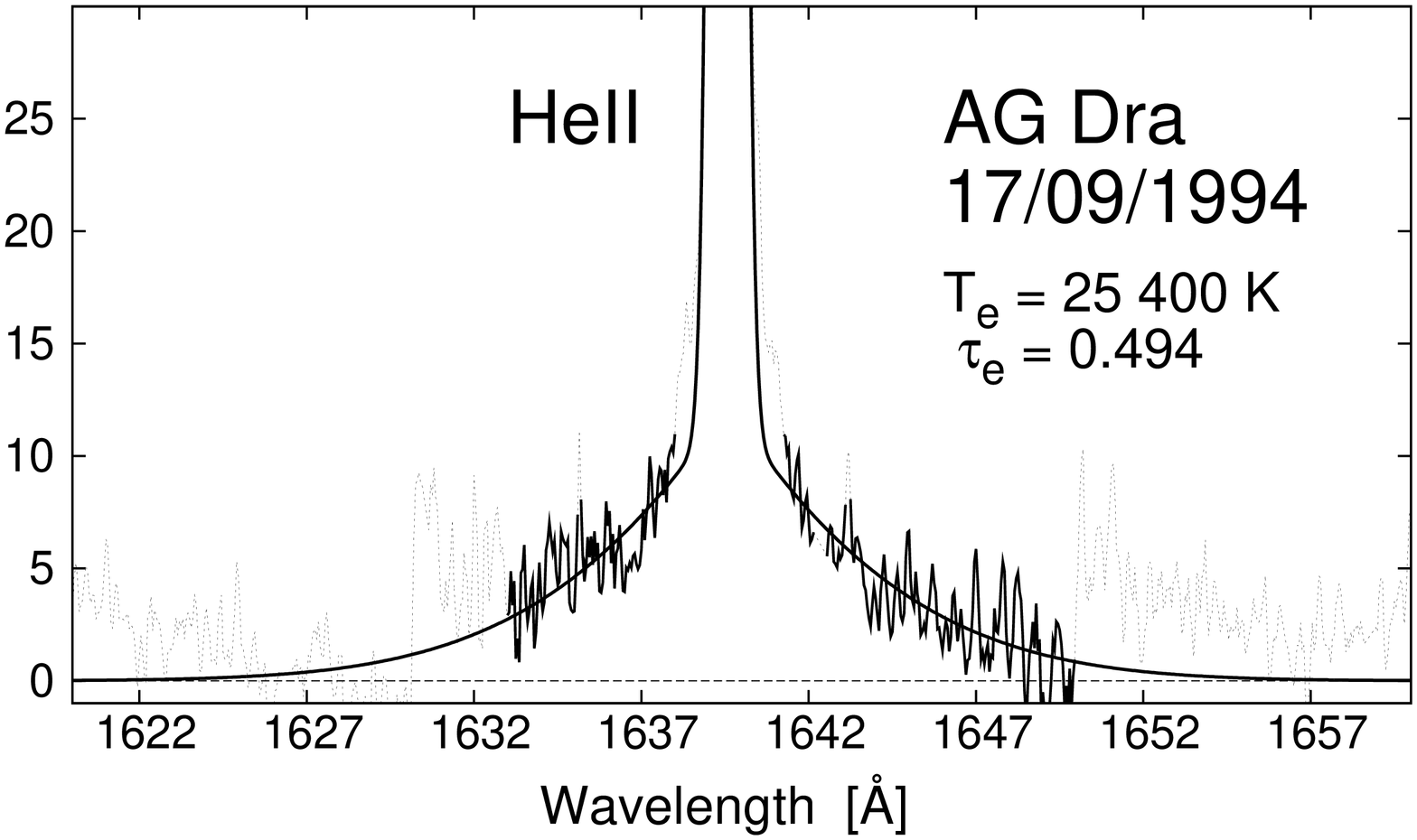}
                      \includegraphics[angle=-90]{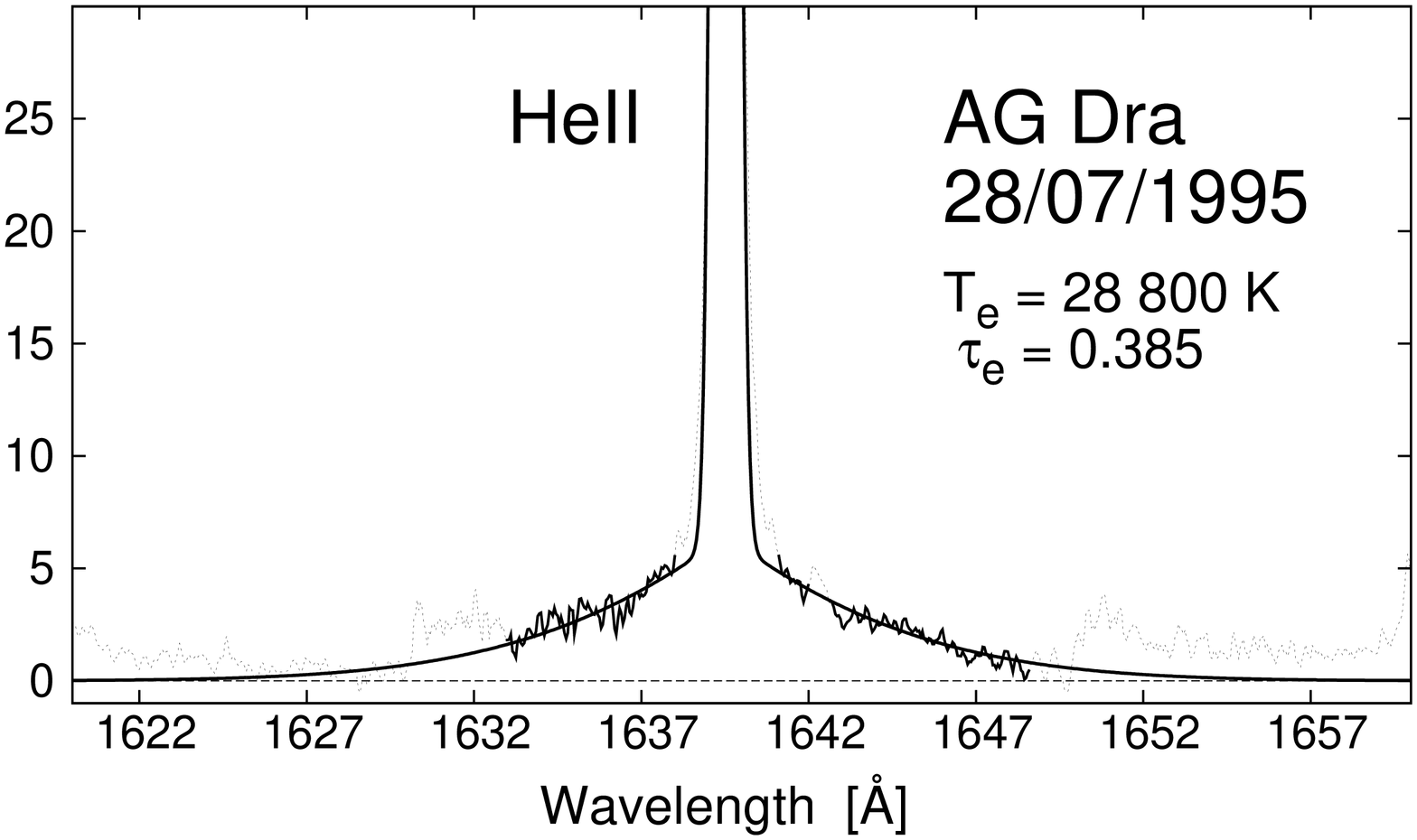}
                      \includegraphics[angle=-90]{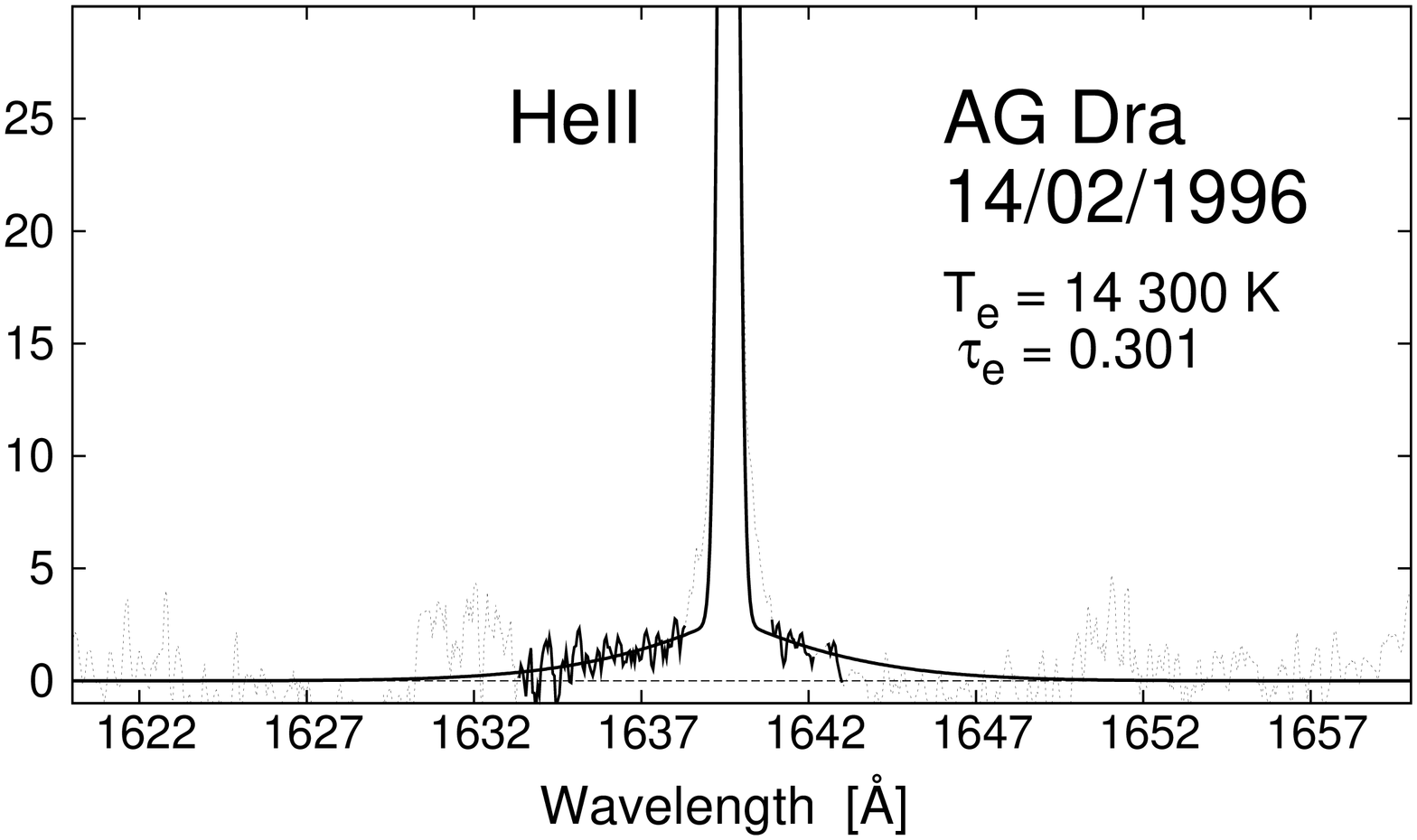}}
\end{center}                               
\caption[]{
Top panel shows the $U$ light curve of AG~Dra from 1977. Active 
phases are characterized by outbursts with multiple maxima. 
The data are from \cite{sk+12}. 
Bottom panels compare the observed (dotted + enhanced line) 
and modelled (solid smooth line) broad wings of the 
O\V\I\ 1032,\,1038\,\AA\ doublet and the He\I\I\ 1640\,\AA\ 
line at different stages of activity. The enhanced parts 
of the observed profile were fitted with the function 
(\ref{eq:psi}). Horizontal dotted line represents the level of 
the continuum. Timing of individual observations are given in 
Table~1 (arrows in the top panel). Fluxes are in units of 
10$^{-12}$\ecsa. 
           }
\label{fig:agmod}
\end{figure*}

\begin{figure*}
\centering
\begin{center}
\resizebox{\hsize}{!}{\includegraphics[angle=-90]{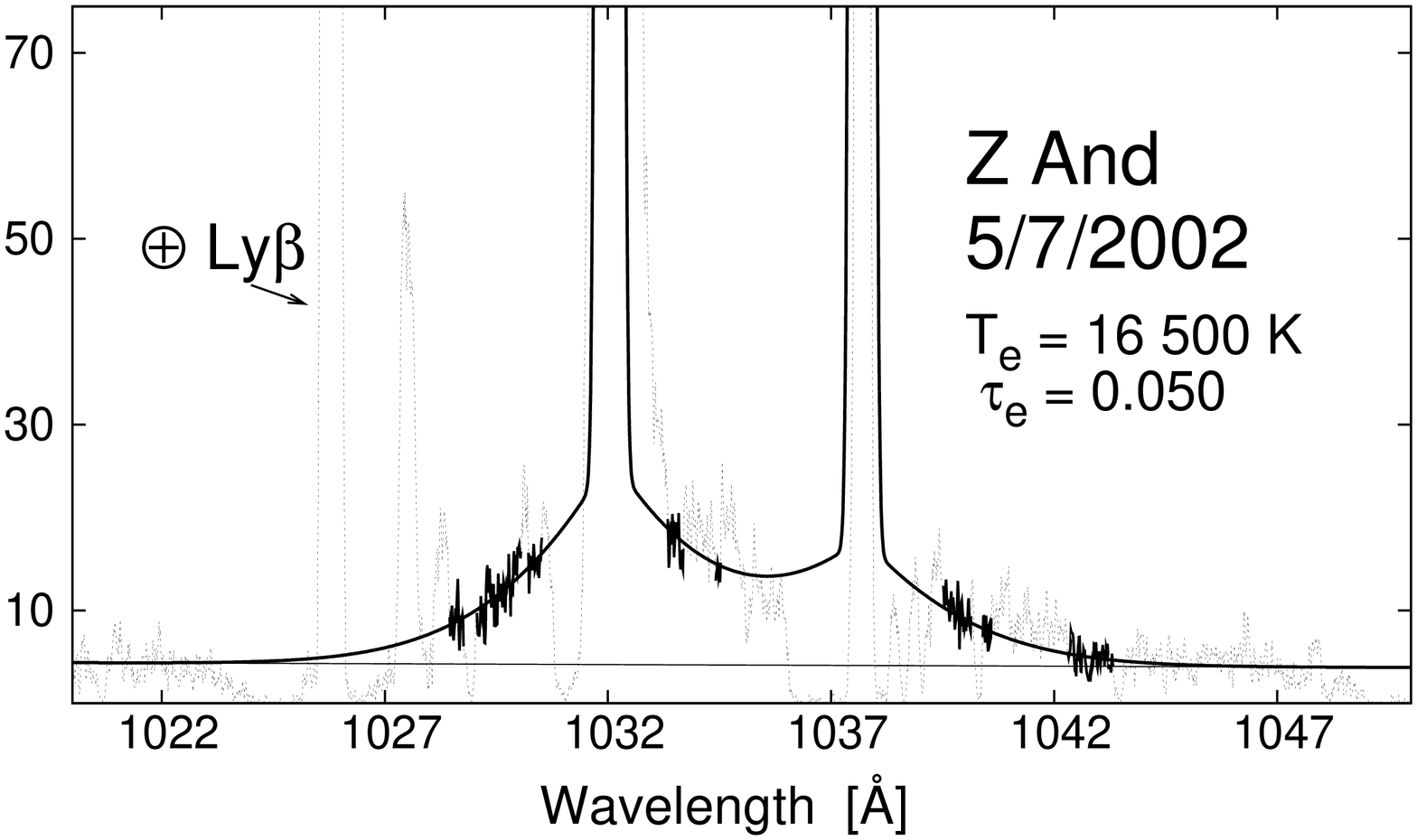}
                      \includegraphics[angle=-90]{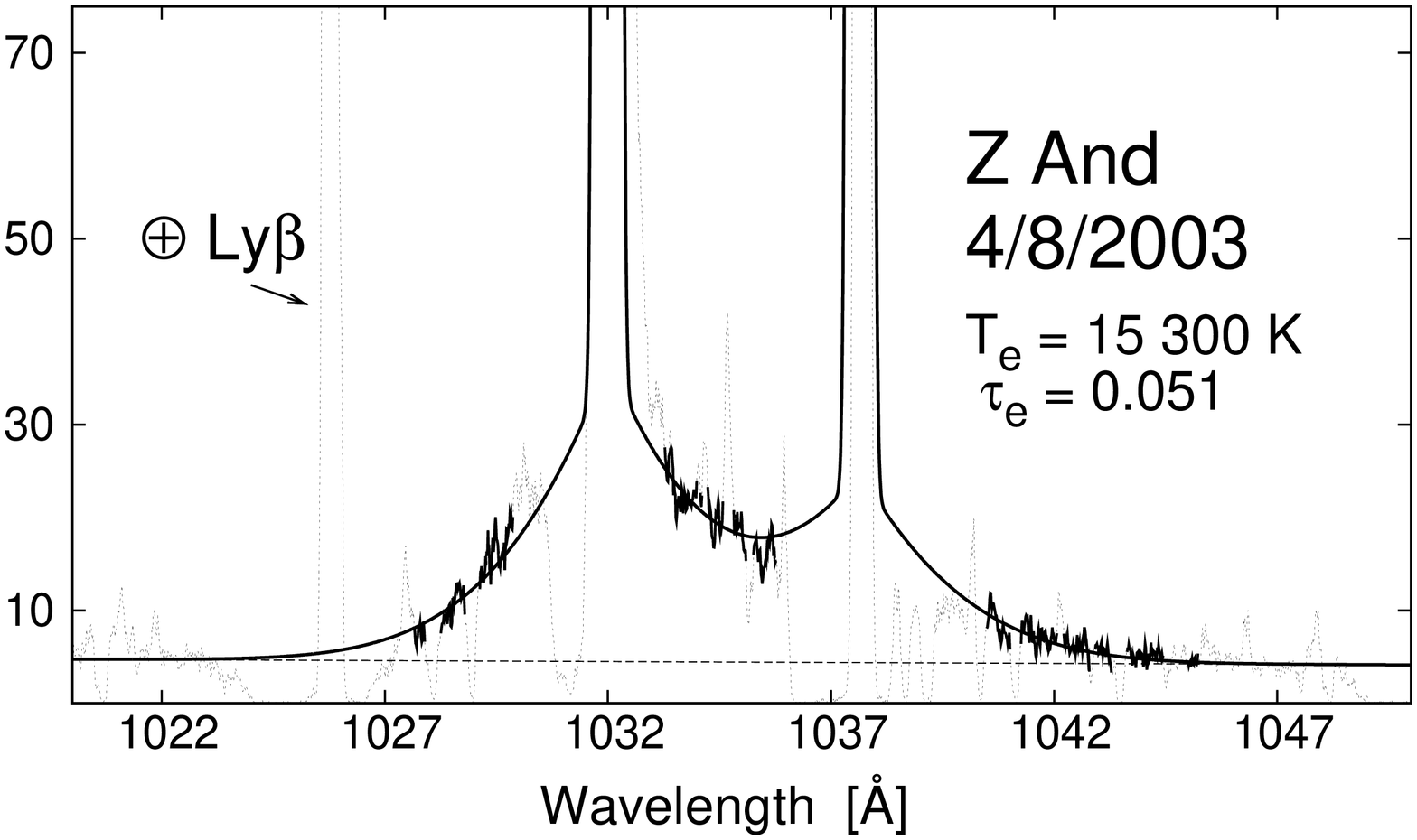}
                      \includegraphics[angle=-90]{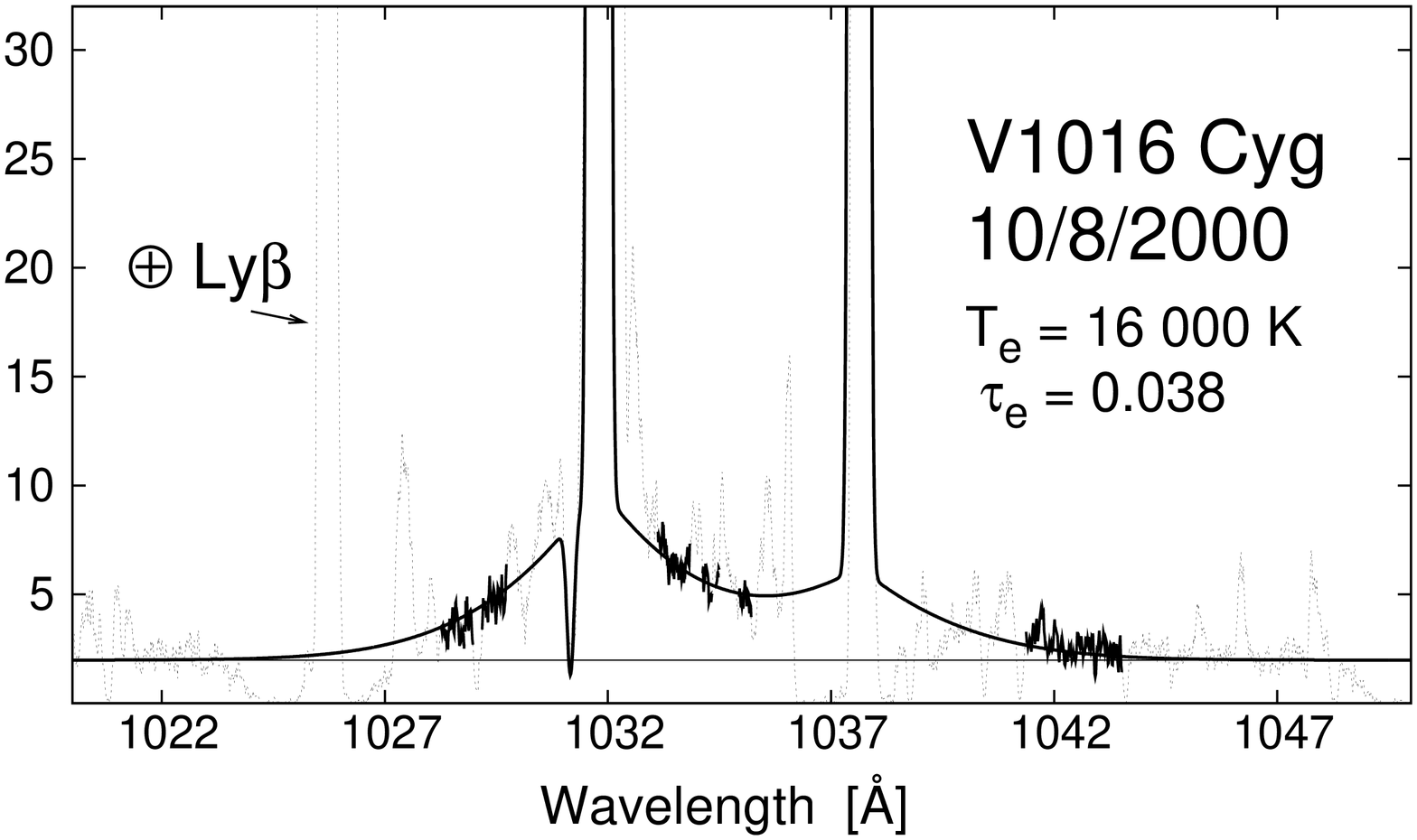}}
\end{center}                               
\caption[]{
As in Fig.~\ref{fig:agmod}, but for Z~And and V1016~Cyg. 
           }
\label{fig:zmod}
\end{figure*}

By modelling the UV/IR continuum of AG~Dra, \cite{sk05} found 
that the mean electron temperature of the nebula during quiescent 
phase and/or small bursts runs between 18\,000 and 21\,800\,K, 
while during major outbursts (1980-81, 1994-5, 2006-7, 
Fig.~\ref{fig:agmod}), the nebula 
significantly strengthens and increases its mean $T_{\rm e}$ 
to $\sim35\,000$\,K. In this work we confirmed these results 
independently by the profile-fitting analysis of the electron 
scattered wings. We found that during a quiescent phase, the 
mean $T_{\rm e} = 21\,700 \pm 3\,600$\,K, while during active 
phases $T_{\rm e} = 32\,300 \pm 2\,000$\,K. Also $\tau_{\rm e}$ 
is a function of the star's activity. Our analysis revealed 
$\tau_{\rm e} = 0.063 \pm 0.007$ and $0.64 \pm 0.11$ during 
quiescence and activity, respectively (Table~2, 
Fig.~\ref{fig:agmod} and \ref{fig:zmod}). 

In the 15/03/2007 spectrum, a relatively high value of 
$\tau_{\rm e} = 0.20$, but very faint wings of the O\V\I\ doublet 
(see Fig.~\ref{fig:agmod}) is caused by the weak incident line 
flux, $F_0$, because $F_{\rm wing}/F_0 = 1 - e^{-\tau_{\rm e}}$ 
(Sect.~3.1., Fig.~\ref{fig:in}). So, the well detectable electron 
scattering wings reflect a relatively large amount of free 
electrons on the line of sight during the transition from 
the major 2006 outburst. 
%
%
%
%

\subsubsection{Z~And}
Z~And is considered to be a prototype of symbiotic stars. Here 
the white dwarf accretes from the stellar wind of a M3-4\,III 
red giant \citep[e.g.][]{fer+88}. During active phases, the 
light curve shows 2--3\,mag brightenings, while the quiescent 
phase is characterized by a wave-like orbitally-related 
variation \citep[e.g.][]{bel85,formiggini94,sk+06}. 

Two \textsl{FUSE} spectra used in this work were observed at 
the end of the major 2000-03 outbursts, 05/07/2002, and 
just after the optical rebrightening on 04/08/2003 
\citep[see Fig.~1 of][]{sk+06}. 
The electron scattering wings observed during both dates 
(Fig.~\ref{fig:zmod}) corresponded to very similar quantities 
of the fitting parameters, 
$T_{\rm e} \sim 16\,500$\,K, $\tau_{\rm e} \sim 0.050$ and 
$T_{\rm e} \sim 15\,300$\,K, $\tau_{\rm e} \sim 0.051$, 
respectively (Table~2). Electron temperatures are consistent 
with those derived from modelling the SED in the continuum 
\citep[][]{sk05}. 
%
%
%
\begin{table*}
\begin{center}
\caption{
Best solutions of Eq.~\eqref{eq:chi2} for $\tau_{\rm e}$ and 
$T_{\rm e}$ and their ranges $\Delta\tau_{\rm e}$ and 
$\Delta T_{\rm e}$, corresponding to 
$\Delta F^{\rm obs}(\lambda_{\rm i})$ (see Sect.~3.3). 
         }
\begin{tabular}{ccccccr}
\hline
\hline
 Date & Stage$^{\star}$ & $\tau_{\rm e}$ & $\Delta\tau_{\rm e}$ 
 & $T_{\rm e}$ & $\Delta T_{\rm e}$ &  $\chi^{2}_{\rm red}$ \\
(dd/mm/yyyy) &  &  &  & [K] & [K] & \\
\hline\\[-3mm]
\multicolumn{7}{c}{AG~Dra} \\
\hline
 11/12/1981 & A & 0.39  & 0.26--0.55 & 40\,100 & 22\,500--47\,200 & 0.6 \\
 18/09/1993 & T & 0.18  & 0.15--0.21 & 20\,200 & 14\,600--23\,100 & 1.9 \\
 29/06/1994 & A & 0.39  & 0.34--0.44 & 32\,700 & 29\,700--34\,900 & 1.6 \\
 09/07/1994 & A & 1.17  & 1.0--1.4   & 31\,000 & 28\,800--34\,500 & 3.6 \\
 12/07/1994 & A & 0.86  & 0.70--1.1  & 29\,100 & 25\,300--38\,700 & 3.0 \\
 28/07/1994 & A & 0.80  & 0.69--0.95 & 41\,300 & 40\,100--46\,700 & 2.5 \\
 17/09/1994 & A & 0.49  & 0.36--0.66 & 25\,400 & 14\,900--27\,400 & 1.2 \\
 28/07/1995 & A & 0.39  & 0.33--0.46 & 28\,800 & 27\,500--33\,500 & 0.9 \\
 14/02/1996 & T & 0.30  & 0.26--0.35 & 14\,300 & 13\,700--15\,800 & 12.1\\
 22/11/1996 & T & 0.23  & 0.13--0.34 & 23\,900 & 18\,600-$>$90\,000 & 11.6\\
 16/03/2000 & Q & 0.044 & 0.033--0.056 & 12\,800 & 7\,600--18\,500& 1.0 \\
 25/04/2001 & Q & 0.065 & 0.051--0.078 & 23\,100 & 22\,900--25\,200& 1.9 \\
 14/11/2003 & T & 0.084 & 0.073--0.096 & 29\,900 & 16\,200--41\,900& 1.2 \\
 24/06/2004 & Q & 0.063 & 0.051--0.074 & 20\,300 & 11\,200--24\,000& 0.8 \\
 25/12/2004 & Q & 0.078 & 0.066--0.091 & 30\,700 & 29\,100--38\,900& 1.8 \\
 15/03/2007 & T & 0.20  & 0.16--0.25   & 21\,300 & 9\,600--27\,900 & 2.3 \\
\hline\\[-3mm]
\multicolumn{7}{c}{Z~And} \\
\hline
 05/07/2002 & Q & 0.050 & 0.042--0.058 & 16\,500 & 8\,600--22\,300 &  2.3 \\
 04/08/2003 & Q & 0.051 & 0.045--0.056 & 15\,300 & 10\,300--22\,400&  1.5 \\ 
\hline\\[-3mm]
\multicolumn{7}{c}{V1016~Cyg} \\
\hline
 10/08/2000 & Q & 0.038 & 0.031--0.044 & 16\,000 & 8\,200--26\,100 &  2.0 \\
\hline
\end{tabular}

$^{\star}$ A -- active phase, Q -- quiescent phase,
           T -- transition to quiescence
\end{center}           
\end{table*}           

\subsubsection{V1016~Cyg}
V1016~Cyg is a member of a small group of symbiotic stars called 
symbiotic novae. In 1964 it underwent a nova-like outburst 
\citep[][]{mccuskey65}, during which the star's brightness 
increased from m$_{\rm pg} \sim 15.5$ to $\sim 10.5$\,mag in 
1971, following a gradual small decrease to $\sim 11.7$ in 2000 
\citep[see Fig.~1 of][]{parimucha+02}. 

There was only one well exposed spectrum in the \textsl{FUSE} 
archive, from 10/08/2000. The wings of the O\V\I\ doublet are 
clearly visible, although the contamination of the 1038\AA\ 
line is significant ($E_{\rm B-V} = 0.28$\,mag). In this 
spectrum we were able to fit also the sharp absorption 
component in the P-Cyg profile of the 1032\AA\ line 
(Fig.~\ref{fig:zmod}). 
Fitting parameters of the nebula, $\tau_{\rm e} = 0.038$ 
and $T_{\rm e} = 16\,000$\,K are similar to those derived 
from observations during quiescent phase of AG~Dra. 
The very small value of $\tau_{\rm e}$ is a result of 
a very weak wings with respect to the strong central 
emission core, $F_{\rm wing}/F_0 = 0.037$. 

\subsection{Thomson scattering during quiescent phase}

Fitting the extended faint wings of the O\V\I\ and He\I\I\ 
lines by Thomson scattering suggested a connection between 
$\tau_{\rm e}$ and the level of the star's activity 
(Fig.~\ref{fig:agmod}). 

\begin{figure}
\centering
\begin{center}
\resizebox{\hsize}{!}{\includegraphics[angle=-90]{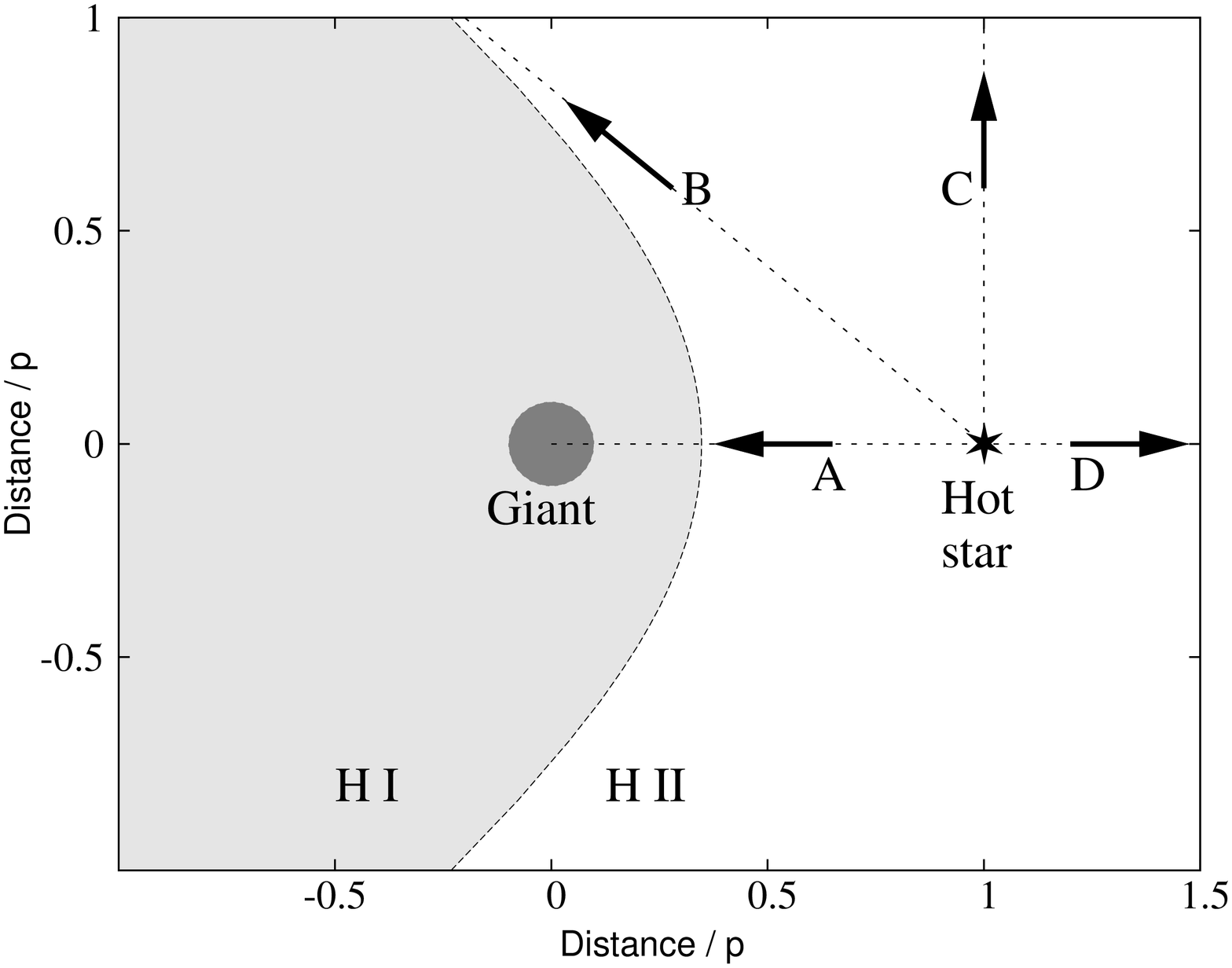}}
\resizebox{\hsize}{!}{\includegraphics[angle=-90]{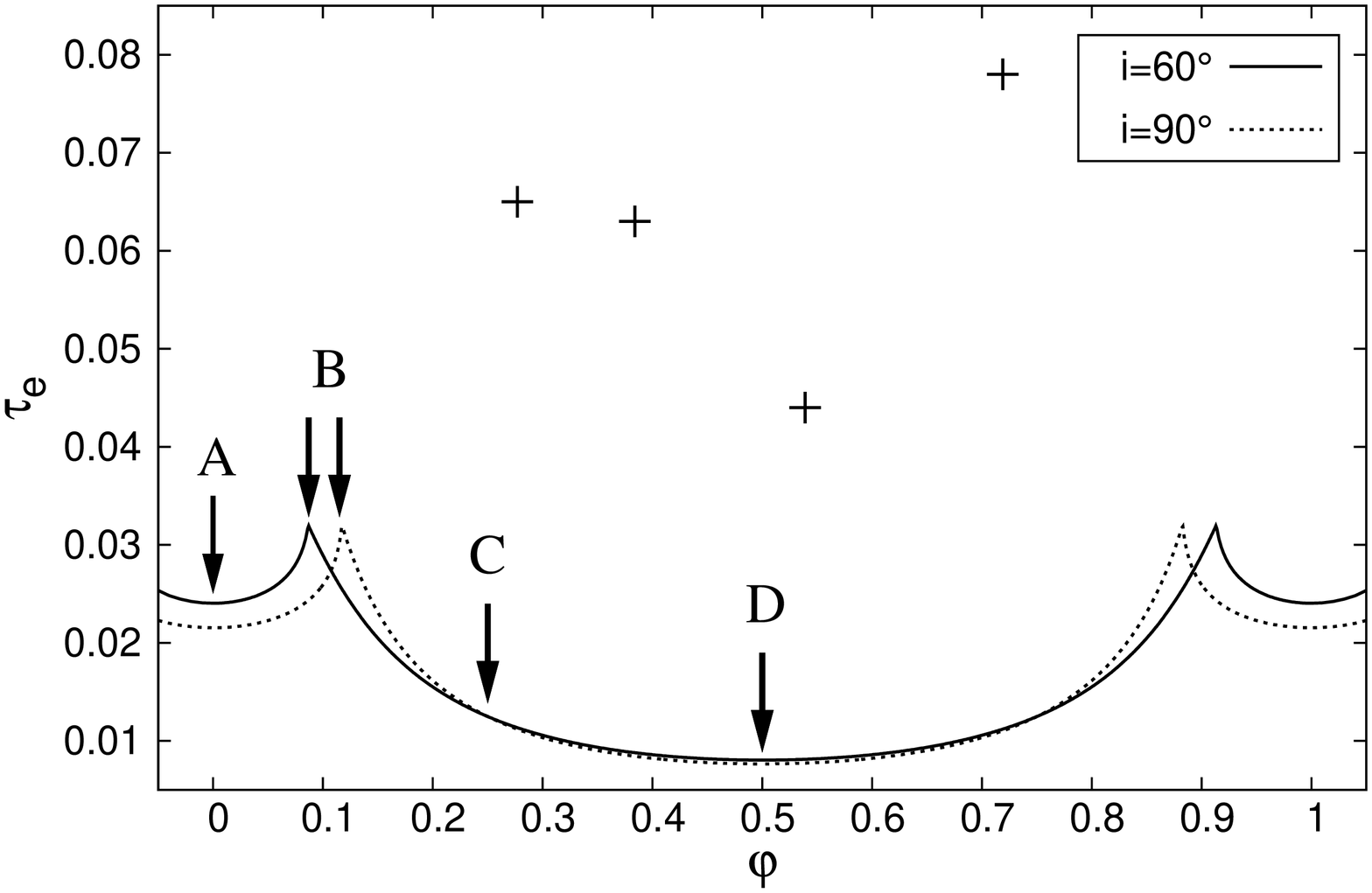}}
\end{center}     
\caption[]{
Top panel shows the ionization structure of AG~Dra as seen 
pole-on. The H\I/H\I\I\ boundary was calculated according to 
Eq.~(\ref{eq:fx0}) and $X = 8$ (Sect.~4.2.). Bottom panel 
displays the corresponding function $\tau_{\rm e}(\varphi)$ 
for the orbital inclination of 60 and 90$^{\circ}$. Labels 
A,B,C,D mark directions at specific orbital phases. Crosses 
correspond to $\tau_{\rm e} = 0.044 - 0.078$, derived 
from the spectra during quiescent phase (Table~2). 
           }
\label{fig:tfi}
\end{figure}

During quiescent phase, it is assumed that the symbiotic nebula 
arises from ionizing a portion of the neutral wind from the giant 
only, i.e. the wind from the hot star is neglected. 
As the source of neutral particles and that of ionizing photons 
are separated, the nebula will be spread asymmetrically around 
the hot star in the binary. This implies that the column densities 
of free electrons on the line of sight to the hot star, and thus 
$\tau_{\rm e}$, will also be a function of the orbital phase, 
$\varphi$. Here we investigate the function $\tau_{\rm e}(\varphi)$ 
for the simplest (idealized) case as outlined by STB. 
In particular, we assume a spherically-symmetric unperturbed 
wind from the giant, whose particles are accelerated along 
the $\beta$-law, and the stationary situation (i.e. no 
binary rotation and no gravitational attraction of the accretor 
to the wind are included). 
Under these assumptions, the extent of the ionized zone during 
quiescence can be obtained from a parametric equation 
\begin{equation}
   X = f(r,\vartheta),
\label{eq:fx0}
\end{equation}
which solution defines the boundary between neutral and ionized 
gas at the orbital plane, determined by a system of polar coordinates, 
$r,\vartheta$, with the origin at the hot star. The function 
$f(r,\vartheta)$ was treated for the first time by STB 
for a steady state situation and pure hydrogen gas. \cite{nv87} 
considered also a contribution of free electrons from singly 
ionized helium, because its zone nearly overlaps that of the 
H\I\I\ region. This increases the electron concentration in 
the ionized zone by a factor of $(1 + a({\rm He}))$, where 
$a({\rm He})$ is the abundance by number of He relative to H. 
Following derivation of \cite{nv87}, but replacing the terminal 
velocity of the wind, $v_{\infty}$, by its $\beta$-law 
distribution 
\begin{equation}
 v_{\rm wind}(r) = v_{\infty}
                \Big(1- \frac {R_{\rm g}}{r}\Big)^{\beta}, 
\label{eq:vwind}
\end{equation}
where $r$ is counted from the centre of the cool giant with 
the radius $R_{\rm g}$ (= the beginning of the wind), 
we can express the terms of Eq.~(\ref{eq:fx0}) as 
\begin{equation}
 X = 
 \frac{4 \pi \mu^2 m_{\rm H}^2}
      {\alpha_{\rm B}({\rm H},T_{\rm e}) (1+a({\rm He}))}p L_{\rm H}
 \left(\frac{v_{\infty}}{\dot M_{\rm g}}\right)^2, 
\label{eq:x}
\end{equation}
and
\begin{equation}
  f(u,\theta) = 
  \int\limits_{0}^{u_{\theta}}
  \frac{u^2}
  {l^{4}\left(1-\frac{R_{\rm g}}{l\times p}\right)^{2\beta}}{\rm d}u. 
\label{eq:futh}
\end{equation}
The acceleration parameter in the wind (\ref{eq:vwind}), 
$\beta = 2.5$ for red giants \citep[][]{sch85}, $\mu$ is the 
mean molecular weight, $m_{\rm H}$ is the mass of the hydrogen 
atom, $\alpha_{\rm B}({\rm H},T_{\rm e})$ stands for the total 
hydrogenic recombination coefficient for case $B$, $p$ is the 
separation of the binary components, $L_{\rm H}$ is the flux 
of hydrogen ionizing photons (s$^{-1}$) and $\dot M_{\rm g}$ 
is the mass-loss rate from the giant. 
Finally, we expressed the radial distances $s$ and $r$ in units 
of $p$, defining $u = s/p$ and 
$l = r/p = \sqrt{u^2 + 1 - 2u\,\cos\theta}$. 
Solutions of Eq.~(\ref{eq:fx0}) for $u_{\theta}(= s_{\theta}/p)$ 
at directions $\theta$ define the ionization H\I/H\I\I\ boundary. 
For the AG~Dra parameters 
\citep[$p = 355$\ro, 
$L_{\rm H} = 3.4\times 10^{46}$\,s$^{-1}$, 
$v_{\infty} = 30$\,\kms, 
$R_{\rm g} = 33$\ro, 
$\dot M_{\rm g} = 3.2\times 10^{-7}$\myr\ and 
$\alpha_{\rm B} = 1.4\times 10^{-13}$\,cm$^{3}$ s$^{-1}$,][]
{mik95,fe00,sk05,sk+09} the ionization parameter 
$X = 8$, which corresponds to an open H\I\I\ zone 
around the hot star (see Fig.~\ref{fig:tfi}). 
%
%

Having defined the ionization structure in the binary, we can 
calculate the function $\tau_{\rm e}(\varphi)$ by integrating 
the electron concentration throughout the ionized zone along 
the line of sight to the hot star, i.e. 
\begin{equation}
  \tau_{\rm e} = \sigma_{\rm T}
  \int\limits_{0}^{s_{\theta}}n_{\rm e}(s){\rm d}s,
\label{eq:tfi}
\end{equation}
where the electron concentration 
$n_{\rm e}(s) = (1 + a({\rm He}))n(r)$ and $n(r)$ is the 
density of hydrogen atoms in the wind from the giant. If 
the line of sight is not passing through the neutral region, 
$s_{\theta} \rightarrow \infty$ (in practice we adopted 
$s_{\theta} = 100\times p$). 
To calculate correctly $\tau_{\rm e}(\varphi)$ for AG~Dra, we 
considered its orbital inclination $i = 60^{\circ}$ 
\citep[][]{ss97}. 
The function $\tau_{\rm e}(\varphi)$ is plotted in 
Fig.~\ref{fig:tfi}. 
A maximum value of $\tau_{\rm e}$ ($\sim$0.03) is around 
$\varphi = 0.1$, when the line of sight passes the ionization 
region as an asymptote to the boundary (the direction B in the 
figure). A minimum of $\sim$0.007 corresponds to the position 
with the hot star in front ($\varphi = 0.5$), because of 
the lowest densities of the wind from the giant. 
Figure~\ref{fig:tfi} demonstrates that theoretical 
$\tau_{\rm e}(\varphi)$ function is significantly below 
the values derived from observations during quiescent phases 
(crosses in the figure, Table~2). 
This implies that the ionized fraction of the unperturbed 
wind from the giant is not capable of producing the observed 
electron scattering wings. 

In a more realistic case the density distribution in a binary 
with mass-losing giant is determined mainly by the rotation 
of the binary and accretion by its compact companion, as was 
demostrated by several hydrodynamical simulations 
\citep[e.g.][]{theus+jor,bisikalo+95,mast+morr,nagae+04}. 
In these studies, it was shown that the regions with highly 
increased density around the accretor, around the mass-losing 
star and the whole binary, and behind the accretor (opposite 
to its orbital motion) were in the form of a disc, a spiralling 
arm and an elongated accretion wake, respectively. 
As a result, the column density of free electrons in the 
direction to the accretor can be enriched by the ionized 
material accumulated at/around the accretion disk at each 
position of the binary. Additional extremes can be expected 
for highly inclined orbits, when the line of sight passes 
throughout a higher density structure 
\citep[see e.g. Fig.~6 of][]{dumm+00}. 
A lower orbital inclination smooths out the density contrasts 
at different phases \citep[see Fig.~2 of][]{theus+jor}. 
Accordingly, in the real case, the faint electron-scattering 
wings during quiescent phases with 
$\tau_{\rm e} = 0.044-0.078$ can be caused mainly by free 
electrons from/around the accretion disk and the ionized 
wind from the hot star, which corresponds to the total column 
density, 
$N_{\rm e} = \tau_{\rm e}/\sigma_{\rm T} \la 10^{23}$\,\cmd. 
The presence of the latter was proved by more authors 
\citep[e.g.][]{vogel93,nsv95,sk06}. 
%
%
\begin{figure}
\centering
\begin{center}
\resizebox{\hsize}{!}{\includegraphics[angle=-90]{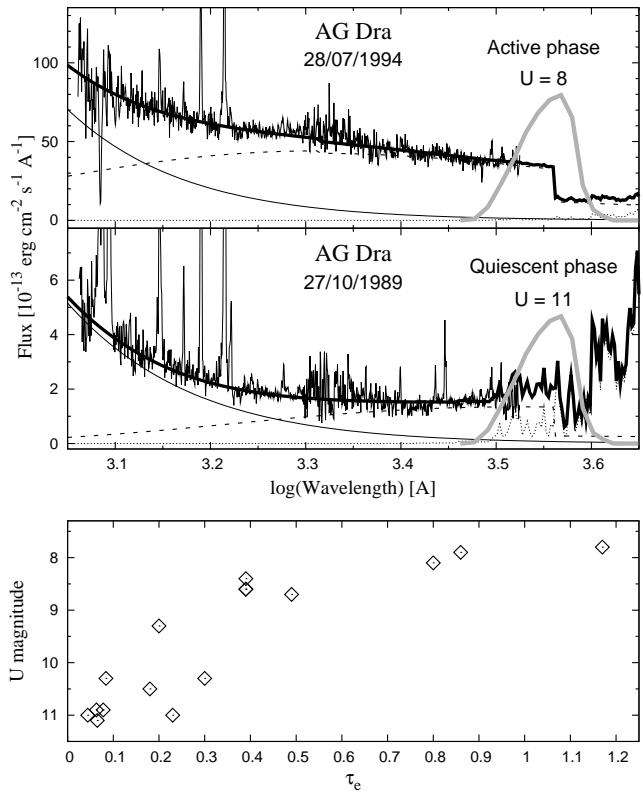}}
\end{center}
\caption[]{
Top panels show the \textsl{IUE} spectra of AG~Dra during 
the 1994 active phase (SWP51632+LWP28752) and a quiescent 
phase (SWP37473+LWP16675). The dashed, solid thin and dotted 
line represent the continuum from the nebula, the hot star 
and the giant, respectively. The solid thick line is their 
sum, the model SED \citep[see][in detail]{sk05}. 
The gray line figures schematically the profile of the 
photometric $U$ filter. Bottom panel shows the dependence 
of $\tau_{\rm e}$ on the $U$-magnitude 
(Fig.~\ref{fig:agmod}, Table~2). 
            }
\label{fig:sed}
\end{figure}   

\subsection{Thomson scattering during active phase}

It is well known that during active phases the hot stars 
in symbiotic binaries enhance significantly the mass-loss 
rate \citep[e.g.][]{fc+95,nsv95,crok+02,sk06}. 
The ejected material is ionized by the luminous central hot 
star, which thus enhances radiation from the symbiotic nebula. 
For example, \cite{sk05} derived a factor of $\approx$10 
stronger nebular emission in the continuum during active 
phases with respect to values measured during quiescence. 
The enhanced mass-loss rate from the hot star thus represents 
a significant supplement of free electrons into the nebula. 
As a result, the electron optical depth, 
 $\tau_{\rm e} = \sigma_{\rm T} N_{\rm e}$, 
will be considerably larger during active phases (Table~2). 
Here, this is well documented by the series of the AG~Dra 
spectra (Fig.~\ref{fig:agmod}). 
The model SED demonstrates that the nebular radiation dominates 
also the spectral region of the photometric $U$ filter 
(see top panels of Fig.~\ref{fig:sed}). 
Therefore, the level of the AG~Dra activity is well mapped 
with the $U$ light curve (top panel of Fig.~\ref{fig:agmod}), 
which thus explains the relationship between $\tau_{\rm e}$ 
and the star's brightness in $U$ (bottom panel of 
Fig.~\ref{fig:sed}). A large scatter around 
$\tau_{\rm e} = 0.2$ is probably connected with the transition 
phase, when the nebula can be partially optically thick 
in the continuum. 

We can conclude that during active phases the large values of 
$\tau_{\rm e} = 0.39 - 1.17$ 
(i.e. $N_{\rm e} = 0.58 - 1.8 \times 10^{24}$\,\cmd) 
are caused by a supplement of free electrons into the binary 
environment as a result of the enhanced mass-loss rate from 
the hot star. 

\section{Conclusions}

We investigated the effect of Thomson scattering of the strong 
emission lines, O\V\I\,1032\,\AA, 1038\,\AA\ and He\I\I\,1640\,\AA, 
observed in the spectra of symbiotic stars AG~Dra, Z~And and 
V1016~Cyg. 
Our models of their profiles are in good agreement with those 
observed by \textsl{FUSE}, \textsl{BEFS}, \textsl{TUES} and 
\textsl{IUE} satellites (Figs.~\ref{fig:agmod} and \ref{fig:zmod}). 
This supports the idea that the broad wings of these lines result 
from the scattering of the line photons on free electrons. Their 
profile is given by two fitting parameters, $\tau_{\rm e}$ and 
$T_{\rm e}$ (Table~2), which characterize the scattering region, 
i.e. the symbiotic nebula. Particular results of our analysis 
can be summarized as follows. 
\begin{enumerate}
\item 
The presence of the electron scattering wings in the line 
profiles of highly ionized elements locates their origin 
mostly to the vicinity of the hot star in the binary, with 
the highest density on the line of sight. 
\item
During quiescent phases, the mean 
   $T_{\rm e} = 19\,200 \pm 2\,300$\,K, 
while during active phases 
   $T_{\rm e} = 32\,300 \pm 2\,000$\,K. 
This findings agrees well with quantities derived independently 
by modelling the SED. 
\item 
The electron optical depth also depends strongly on the star's 
activity. During quiescent phases, 
  $\tau_{\rm e} = 0.056 \pm 0.006$, 
while during active phases, 
  $\tau_{\rm e} = 0.64 \pm 0.11$ 
(Table~2, Figs.~\ref{fig:agmod} and \ref{fig:zmod}). 
Uncertainties in the results given in (ii) and (iii) 
represent rms errors of the average values. 
\item
During quiescent phases, the ionized fraction of the wind 
from the giant within a simple (STB) model (Sect.~4.2) is 
not capable of giving rise a measurable effect of the Thomson 
scattering. In the real case, the faint electron-scattering 
wings with $\tau_{\rm e} = 0.044-0.078$ are caused mainly 
by free electrons from/around the accretion disk and the 
ionized wind from the hot star with the total column density 
of a few $\times 10^{22}$\,\cmd\ 
(Sect.~4.2., Fig.~\ref{fig:tfi}). 
\item
During active phases, the large values of $\tau_{\rm e}$ are 
caused by a supplement of free electrons into the nebula 
from the enhanced wind of the hot star, which increases 
$N_{\rm e}$ to $\sim 10^{24}$\,\cmd\ 
(Sect.~4.3., Fig.~\ref{fig:sed}). 
\end{enumerate}
The presumable relationship between $\tau_{\rm e}$ and the 
level of the activity, as suggested by our profile-fitting 
analysis, could be used in probing the mass-loss rate from 
the hot stars in symbiotic binaries. 

\section*{Acknowledgments}
The authors thank the anonymous referee for constructive comments. 
The spectra used in this paper were obtained from the satellite 
archives with the aid of MAST. STScI is operated by the Association 
of Universities for Research in Astronomy, Inc., under NASA 
contract NAS5-26555. Support for MAST for non-HST data is provided 
by the NASA Office of Space Science via grant NNX09AF08G and by 
other grants and contracts.
The research presented in this paper was in part supported 
by the Slovak Academy of Sciences under a grant 
VEGA No.~2/0038/10, and by the realization of the Project
ITMS No. 26220120029, based on the supporting operational 
Research and development program financed from the European 
Regional Development Fund.

\end{document}